\documentclass[aps,twocolumn,showpacs,preprintnumbers,amsmath,amssymb,nofootinbib,superscriptaddress,showkeys]{revtex4}

\usepackage{hyperref}
\usepackage{float}
\usepackage{epsfig}
\usepackage{graphicx}
\usepackage{mathptmx}      
\usepackage{latexsym}
\usepackage{longtable}
\usepackage{dcolumn}
\usepackage[utf8]{inputenc}

\usepackage{graphicx,color}
\usepackage{amsmath,amssymb,bm}
\usepackage{dcolumn}

\newcommand{\EP}{\epsilon}

\def\XXint#1#2#3{{\setbox0=\hbox{$#1{#2#3}{\int}$}
     \vcenter{\hbox{$#2#3$}}\kern-.5\wd0}}

\def\1{\'{\i}}

\begin{document}

\title{The Low energy structure of the Nucleon-Nucleon
  interaction:\\ Statistical vs Systematic Uncertainties}

\author{R. Navarro P\'erez}\email{navarroperez1@llnl.gov}
\affiliation{Nuclear and Chemical Science Division, Lawrence Livermore National Laboratory, Livermore, CA 94551, USA}
\author{J.E. Amaro}\email{amaro@ugr.es} \affiliation{Departamento de
  F\'{\i}sica At\'omica, Molecular y Nuclear \\ and Instituto Carlos I
  de F{\'\i}sica Te\'orica y Computacional \\ Universidad de Granada,
  E-18071 Granada, Spain.}
  
\author{E. Ruiz
  Arriola}\email{earriola@ugr.es} \affiliation{Departamento de
  F\'{\i}sica At\'omica, Molecular y Nuclear \\ and Instituto Carlos I
  de F{\'\i}sica Te\'orica y Computacional \\ Universidad de Granada,
  E-18071 Granada, Spain.} 

\date{\today}

\begin{abstract}
\rule{0ex}{3ex} We analyze the low energy NN interaction by
confronting statistical vs systematic uncertainties. This is carried
out with the help of model potentials fitted to the Granada-2013
database where a statistically meaningful partial wave analysis
comprising a total of $6713$ np and pp published scattering data from
1950 till 2013 below $350 {\rm MeV}$ has been made.  We extract
threshold parameters uncertainties from the coupled channel effective
range expansion up to $j \le 5$.  We find that for threshold
parameters systematic uncertainties are generally at least an order of
magnitude larger than statistical uncertainties. Similar results are
found for np phase-shifts and amplitude parameters.
\end{abstract}


\maketitle


\section{Introduction}

The goal of the present paper is to quantify the uncertainties on the
knowledge of the NN system from the present available experimental pp
and np scattering data and their uncertainties. We do this from a
comprehensive partial wave analysis (PWA) containing the largest NN
database to date wich permits a statistically self-consistent least
squares fit. From the determination of several statistically
equivalent interactions we deduce the residual systematic differences
in many two body quantities of interest. The main result is the
dominance of these systematic errors over the statistical ones
determined from each interaction separately.

\subsection{Uncertainties in the Nuclear Force}

The NN interaction, as a key building block of nuclear physics, has
traditionally been inferred from pp and np scattering data. This task
is hampered both by the fragmentary body of experiments as well as by
the incomplete status of the models used to analyze them. These
aspects have important consequences regarding the predictive power,
accuracy and precision in theoretical nuclear physics. In particular,
{\it ab initio} calculations of nuclear structure and nuclear
reactions in terms of protons and neutrons as elementary constituents
require the design of Nucleon-Nucleon potentials validated with the
existing scattering information (see
e.g. Ref.~\cite{signell1995nuclear} for a lucid presentation). As it
is well known the form and representation of potentials is not unique,
and the historic evolution reflects this large
diversity~(see~\cite{Machleidt:1992uz} and references therein for a
pre-nineties review). Even if a set of potentials are succesfully
validated against the existing scattering data by an statistically
acceptable $\chi^2$ fit value, the inferred predictions of unmeasured
scattering quantities or other observable quantities such as nuclear
bindings have two sources of uncertainties.  Besides the obvious ones
stemming from the experimental data and whose statistical nature
requires passing elementary statistical tests, one also has a
dependence on the particular choice of potential used to make the fit.
This residual non-statistical dependence is our definition of the
systematic uncertainty. In what follows we will ellaborate on what we
think are elementary aspects of error
analysis~\cite{evans2004probability}, as applied to the NN interaction
since these basic principles can be easily implemented in large
  scale fits and calculations.

The statistical uncertainties are easier to quantify, provided one can
credibly establish that the discrepancies between theory and
experiment are fluctuations whose probability distribution is either
known {\it a priori} or confidently tested {\it a posteriori}. For the
conventional least squares $\chi^2$-fit procedure this corresponds to
test the normality of fitting residuals obtained by building the
differences from the optimized theory and the fitted experimental
data. When this is the fortunate case, the fit is self-consistent as
the {\it a priori} assumption is verified {\it a posteriori} by the
actual fit, and statistical error propagation can routinely be
undertaken. We stress this essential point as it has
too often been ignored in the design of the so-called high-quality
potentials in the past.

Systematic uncertainties are notoriously more difficult to pin down in
general. In fact, there are many ways to quantify the systematic
uncertainties and none of them can be complete within the present
context for two main reasons. On the one hand there are many
imaginable forms of potentials which could fit the existing finite
amount of data with equal statistically meaningfull quality, and thus
in general only lower bounds on the systematics can be estimated. On
the other hand there exist many derived quantities such as scattering
amplitudes, phase-shifts or nuclear binding energies which will
reflect the effect of the systematic error differently on a
quantitative level. This has been the traditional approach to
  systematic uncertainties in the past by trying out different high
  quality potentials in nuclear structure ab initio calculations (see
  e.g.
  \cite{Maris:2008ax,Coraggio:2008in,Bogner:2009bt,Roth:2010bm,Leidemann:2012hr,Marcucci:2015rca,Navratil:2016ycn}).

\subsection{Our contribution to NN analyses}

Our work can be framed within the currently growing efforts to
realistically pin down the existing uncertainties stemming from
different sources in theoretical nuclear
physics~\cite{dudek2013predictive,Dobaczewski:2014jga}. In fact, a
special issue dedicated to this topic appeared recently in
Jour. Phys. G~\cite{JPG-2015}~\footnote{An editorial recommendation on
  the necessity of including uncertainties in theoretical evaluations
  in Atomic and Molecular Physics has been published in
  2011~\cite{PhysRevA.83.040001}.}.  In this paper we will restrict
the analysis to the NN scattering amplitude and try to set lower
bounds both on statistical as well as systematic uncertainties in
terms of a finite number of potentials suitable for nuclear structure
calculations.  The set of model potentials used below is a
convenient tool with the purpose of quantifying the uncertainties.
This is a necessary but already insightful step before extending these
uncertainties to {\it ab initio} nuclear structure calculations. Since
it is naturally assumed that for light nuclei nuclear binding is
mainly sensitive to low energy NN scattering, we also extensively
study the long wavelength limit because also potential model details
are expected to become least relevant. We find that {\it even} in the
low energy limit the systematic uncertainties dominate over those
statistical uncertainties arising directly from the same experimental
data. Our analysis is based on a comparison of 6 statistically
acceptable but different model potentials, i.e., with $\chi^2/{\rm
  d.o.f} \sim 1$ developed by our group fitting the {\it same} self
consistent Granada database comprising a total of 6713 NN
scattering~\cite{Perez:2013jpa}. The present paper is complementary
to~\cite{Perez:2013jpa} and further dedicated studies along these
lines~\cite{Perez:2014yla,Perez:2014kpa}.

\subsection{The NN error analysis in retrospect}

In order to understand our most unexpected result and to provide a
proper perspective, we provide at this point a sucint review with the
historic benchmarks as guidelines highlighting those aspects
specifically dealing with our work. We recommend the comprehensive
presentation covering up to 1992 for a wider
scope~\cite{Machleidt:1992uz}.

The low energy structure of the NN interaction has received a
recurrent attention since the late 40's when Bethe proposed the
effective range expansion~\cite{Bethe:1949yr} (ERE).  The shape and model
independence of the amplitude captured with a few number of parameters
the essence of the NN force; an unique and particularly appealing
universal pattern in the long wavelength limit of short range
interactions. Because this is a low energy expansion of the full
scattering amplitude in powers of small momentum, higher partial waves
are not needed in principle, and one may imagine an ideal situation
with a direct determination from very low energy scattering
only. However, these very low energy data are scarce and an
extrapolation to zero energy must always be made.  Much higher
accuracy can be obtained by intertwining lower and higher energies via
a large scale Partial Wave Analysis (PWA) so that many more data
contribute to the threshold parameters precision when the resulting
scattering amplidues are evaluated at zero energy. While this
procedure largely increases the statistics, this interrelation cannot
be achieved for free.  As we discuss next some unavoidable model
dependence is introduced, thus generating a source of systematic
errors beyond the genuine statistical errors of the PWA.

Indeed, the NN scattering amplitude contains 10 functions of energy
and angle and a complete set of experiments is needed to determine it
without model dependence~\cite{puzikov1957construction}. While the
usefulness of polarization was soon
realized~\cite{schumacher1961usefulness} as well as the strong
unitarity constraints on the uniqueness of the
solution~\cite{alvarez1973complete,Bystricky:1976jr} (see
~\cite{kamada2011determination} for an analytical solution), complete
sets of observables are scarce at the energies relevant to nuclear
structure applications, corresponding to energies below or about pion
production threshold (see also \cite{Arndt:1973ec}). Following the
standard custom, we will take the maximal LAB energy in our potential
analysis to be $350 {\rm MeV}$, which has been the canonical choice
for NN potential fits. As a consequence a PWA in conjunction with the
standard least squares $\chi^2$-method pioneered by Stapp and
Ypsilantis~\cite{Stapp:1956mz} is usually pursued to fit the set of
measured scattering observables at given {\it discrete} energy and
angle values. Soon thereafter, the nowadays widely accepted
probabilistic interpretation of the $\chi^2$-fit in terms of the
p-value was introduced as a measure of the confidence
level~\cite{Cziffra:1959zza,MacGregor:1959zz} (see particularly the
figure in \cite{MacGregor:1959zz} where the p-value is explicitly
displayed). The method was popularized during the 60's by MacGregor
and Arndt with additional implementations, such as a rejection
criterium for the growing number of incompatible
parameters~\cite{arndt1966determination,arndt1966chi,MacGregor:1968zzd}.
This is nothing but the long established Chauvenet's $3\sigma$
criterion (see \cite{taylor}) to statistically reject data sets with
an improbably high or improbably low $\chi^2$ value.

This incomplete and discretized experimental information requires
using smooth interpolating energy dependent functions for the
scattering matrix in all partial waves for nearby but unmeasured
kinematic regions. Alternatively, quantum mechanical potentials with
proper long distance behavior generate analytical energy dependence
with the adequate cut-structure in the complex energy plane (see e.g.
Ref.~\cite{signell1995nuclear} for a review). We follow here the
potential approach since it has the obvious advantage over a mere
partial wave analysis of being of direct use in nuclear structure
calculations, and hence it allows to direct transporting
uncertainties from the data to energy bindings. The potential approach
is subjected to inverse scattering off-shell
ambiguities~\cite{chadan2011inverse} as the potential contains
generically 10 matrix functions~\cite{okubo1958velocity}, manifesting
themselves as a systematic uncertainty in the observables at the
interpolated, not directly measured, energy values. If one fixes a
maximum energy for the PWA the ambiguities reflect the finite spatial
resolution corresponding to the shortest de Broglie wavelength below
which the interaction is not determined by the data. Thus, we might
expect that in the long wavelength limit systematic uncertainties will
be greatly reduced.

\subsection{Organization of the paper}

The paper is organized as follows. In Section~\ref{sec:dat-mod} we
describe the main ideas behind our analysis, reviewing our previous
works as well as a summary of the 6 model potentials used in our
analysis. We remind the effective range expansion for the deuteron
channel in Section~\ref{sec:low-exp}. The convenient and accurate tool
which we will be using to evaluate low energy parameters in the
discrete version of the coupled channel variable S-matrix approach is
presented in Section~\ref{sec:discrete-S}. This framework proves
extremely convenient to discuss the pertinent sampling of the
interaction, and the difference between fine and coarse graining is
addressed in Section~\ref{sec:fine-coarse}.  There, peculiar numerical
aspects of these calculations are also analyzed. Our main numerical
results concerning the comparison between statistical vs systematic
uncertainties are presented in Section~\ref{sec:stat-sys} in terms of
phase-shifts, scattering amplitudes and potentials.  In
Section~\ref{sec:port} we ponder on the portability of phase shift
analyses based on our own experience with nuclear potentials.
Finally, in Section~\ref{sec:conclusions} we summarize our main
results and conclusions. In the appendices we provide some details
concerning three new potentials introduced in the present work.

\section{NN Data, Models and Uncertainties}
\label{sec:dat-mod}

\begin{table*}
 \footnotesize
 \caption{\label{tab:PotentialsSummary} Model Potentials Summary.}
 \begin{ruledtabular}
 \begin{tabular*}{\textwidth}{@{\extracolsep{\fill}} l *9{c}}
   Potential       & Number of Parameters & $N_{np}$ & $N_{pp}$ & $\chi^2_{np}$ & $\chi^2_{pp}$ & $\chi^2/{\rm d.o.f.}$ & p-value  & Gaussianity & 
 Birge Factor \\
   \hline
   DS-OPE          & 46         & 2996     & 3717     & 3051.64       & 3958.08    & 1.05     & 0.32  &  Yes &  1.03   \\
   DS-$\chi$TPE    & 33         & 2996     & 3716     & 3177.43       & 4058.28    & 1.08     & 0.50  & Yes  &  1.04   \\
   DS-$\Delta$BO   & 31         & 3001     & 3718     & 3396.67       & 4076.43    & 1.12     & 0.24  & Yes  &  1.06   \\
   Gauss-OPE       & 42         & 2995     & 3717     & 3115.16       & 4048.35    & 1.07     & 0.33  & Yes  &  1.04   \\
   Gauss-$\chi$TPE & 31         & 2995     & 3717     & 3177.22       & 4135.02    & 1.09     & 0.23  & Yes  &  1.05  \\
   Gauss-$\Delta$BO& 30         & 2995     & 3717     & 3349.89       & 4277.58    & 1.14     & 0.20  & Yes  &  1.07     
 \end{tabular*}
 \end{ruledtabular}
\label{tab:gr-pots}
\end{table*}

\begin{figure}
\centering
\includegraphics[width=\linewidth]{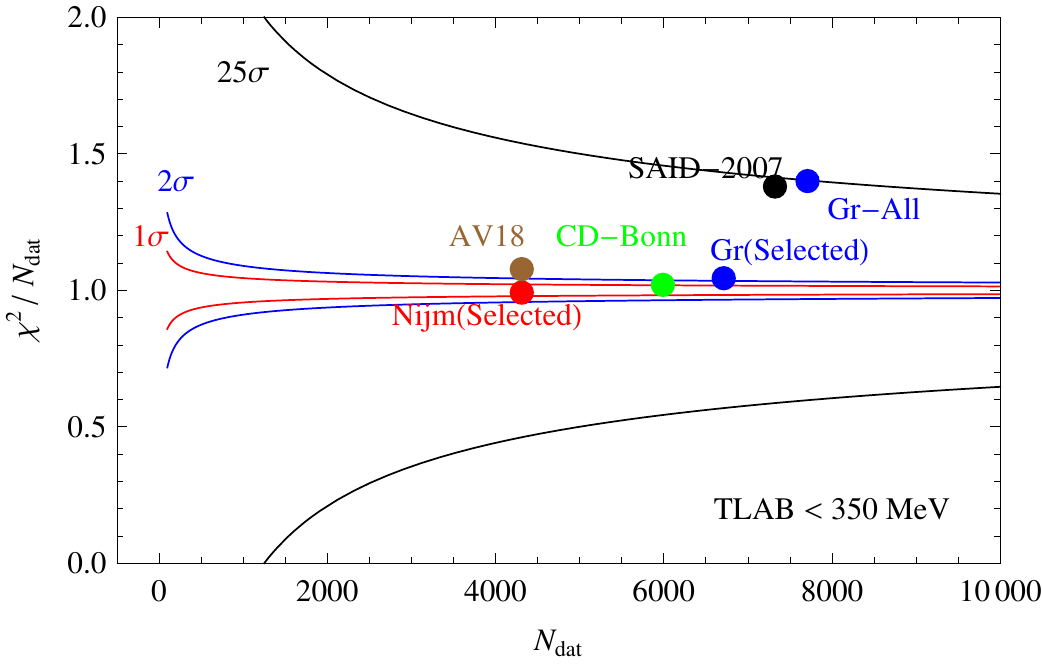}
\caption{Values of the $\chi^2/N_{\rm dat}$ as a function of the
  number of data $N_{\rm dat}$ provided by the experimentalists for
  several fits at LAB energies below $350$MeV. We include the Nijmegen
  PWA~\cite{Stoks:1993tb}, the AV18 potential~\cite{Wiringa:1994wb}
  the CD-Bonn potential~\cite{Machleidt:2000ge}, the Granada
  PWA~\cite{Perez:2013mwa} and the SAID PWA~\cite{SAID}.  In the
  Granada analysis we distinguish between a fit to all the published
  data (Gr-All) and the $3\sigma$ self-consistent database,
  Gr(Selected). For comparison we also plot the $1\sigma$, $2\sigma$
  and $25\sigma$ confidence levels.}
\label{fig:chi2}       
\end{figure}

\subsection{The situation after the Nijmegen-1993 analysis}

The description of NN scattering data by phenomenological potentials
started in the mid-fifties~\cite{Stapp:1956mz} and has been pursued
ever since (see~\cite{Machleidt:1992uz} and references therein for a
pre-nineties review). However a successful fit, determined by the
merit figure $\chi^2/{\rm d.o.f} \sim 1$, was not achieved until 1993
when the Nijmegen group applied in this context Chauvenet's $3\sigma$
rejection criterion already proposed in 1968~\cite{MacGregor:1968zzd}
to statistically discard data sets with an improbably high or
improbably low $\chi^2$ value ~\cite{Stoks:1993tb}. 

In Fig.~\ref{fig:chi2} we illustrate the situation for $\chi^2/N_{\rm
  dat}$.  The number of data is so large that the
$\chi^2$-distribution behaves as a normal distribution with mean value
$\nu $ and variance $2 \nu$~\cite{taylor}. In addition, $\nu \simeq
N_{\rm dat}$, since the number of fitting parameters is usually much
less than $N_{\rm dat}$.  Thus, the mean and variance of the reduced
distribution, $\chi^2/\nu$, are $1$ and $\sqrt{2/\nu}$
respectively. Therefore those fits in Fig.~\ref{fig:chi2} falling
outside the interval $1\pm N \sqrt{2/N_{\rm dat}}$ contain gross
systematic errors with a $68\%,95\%$ confidence level for $N=1,2$
respectively.  For instance, for the SAID database~\cite{SAID} the
total number of data is $N_{\rm dat}= 3075_{pp} + 4159_{np}= 7234$ and
the total $\chi^2 = 4043_{pp} + 6160_{np}=10203$, giving $
\chi^2/N_{\rm dat} = 1.4157$. Then, $\sigma=
\sqrt{2/7234}=0.017$. Thus, the SAID reduced $\chi^2$-value is $ 1.41
= 1 + 25 \sigma$. This value is probably so large because the full
database has been used before the data selection.  
More recent
  chiral motivated interactions usually fit data up to lower energies
  \cite{Gezerlis:2014zia,Entem:2003ft} and are therefore in this
  picture. However, some of their merits have been discussed on
  previous publications \cite{Perez:2014bua,Perez:2016vzj}.

After this first statistically satisfactory Nijmegen study, several
potentials describing data up to a laboratory frame energy of $350$
MeV were developed with similar $\chi^2/{\rm d.o.f}$ values. All of
them include the distinguished charge dependent (CD) one pion exchange
(OPE), magnetic moments, vacuum polarization and relativistic effects
as the long range part of the interaction, and around $40$ parameters
for the short and intermediate range regions \cite{Stoks:1993tb,
  Stoks:1994wp, Wiringa:1994wb, Machleidt:2000ge,Gross:2008ps}. These
OPE-tailed potentials with $\chi^2/{\rm d.o.f} \lesssim 1$ have played
a major role in nuclear physics. It should be noted, however, that an
error analysis of these potentials based on the finite experimental
scattering accuracy has been overlooked, and as a consequence any {\it
  ab initio} nuclear structure calculations using them are unable to
quantify the impact of NN scattering uncertainties in nuclear
bindings. 

\subsection{Granada-2013 database and Potentials}

In a recent paper we have updated the Nijmegen PWA by including data
up to 2013, improving after \cite{Gross:2008ps} the $3\sigma$
criterion to select a self-consistent database with $N=6713$ np and pp
scattering data and providing statistical error bars to the fitting
parameters~\cite{Perez:2013mwa, Perez:2013jpa}. The self-consistent
Granada database is available for download~\cite{GranadaDB}.  The
delta-shell (DS) representation of the potential allowed the
propagation of statistical uncertainties from the scattering data into
potential parameters, phase shifts, scattering amplitudes and deuteron
properties. This was possible due to the simplification in calculating
the Hessian matrix.  Subsequently, we have extended the DS potential
including chiral two pion exchange ($\chi$TPE) in the intermediate and
long range regions~\cite{Perez:2013oba, Perez:2013cza}.  We also
introduced a local, smooth potential parameterized as a sum of
Gaussians (SOG) with OPE~\cite{Perez:2014yla}. In
appendices~\ref{app:GausTPE} and \ref{app:DeltaBO} we introduce three
new model interactions.  One of them is a SOG model with
$\chi$TPE. The other two contain $\Delta$ resonances as dynamical
degrees of freedom through the Born-Oppenheimer approximation
($\Delta$BO), and are modeled either by DS or SOG at short
distances. Note that only the SOG model potentials are smooth and can
be plotted. The statistical features of the 6 model interactions are
displayed in Table~\ref{tab:gr-pots} where we provide the number of
Parameters for np and pp scattering, , $N_{np}$ and $N_{pp}$
respectively, as well as the total $\chi^2$ values in each separate
case. The p-value corresponds to the actual $\chi_{\rm min}^2/\nu$
value. As we have stressed in our previous
works~\cite{Perez:2014yla,Perez:2014kpa} one can globally sligthtly
enlarge the experimental uncertainties by the so-called Birge factor
provided the residuals pass a gaussianity test. After this re-scaling
the p-value becomes 0.68 for a $1\sigma$ confidence level and hence
all potentials become statistically equivalent. As can be seen all of
our potentials incorporate the appropriate propagation of statistical
uncertainties. This has been possible because the residuals of our
fits are normally distributed.  This requirement of the $\chi^2$
method, has been verified {\it a posteriori} with a high confidence
level.  We stress that a lack of normality in the residuals would
strongly suggest the presence of systematic uncertainties in the
analysis, disallowing the propagation of statistical errors. The same
database has also been recently used to fit a chiral TPE potential
that directly includes delta excitation~\cite{Piarulli:2014bda}.  In
the cases where normality is unequivocally fulfilled we have
propagated statistical uncertainties by applying the bootstrap Monte
Carlo method {\it directly} to the experimental
data~\cite{Perez:2014jsa}, which simulates an ensemble of conceivable
experiments based on the experimental uncertainties estimates. A
similar method has successfully been applied to estimate the
statistical uncertainty in the triton binding energy solving the
Faddeev equations~\cite{Perez:2014laa} and the alpha-particle using
shell model techniques~\cite{Perez:2015bqa}. A similar propagation to
the $\alpha$-particle binding energy solving the Faddeev-Yakubovsky
equations~\cite{Perez:2016oia} has been advanced recently.

In Fig.~\ref{fig:chi2} the $N\sigma$ confidence level corresponds to
the interval $\chi^2 /N_{\rm dat}=1\pm N\sqrt{2/N_{\rm dat}}$ and the
corresponding p-value is $p=1-\int_{-N}^{N} dx e^{-x^2/2}
/\sqrt{2\pi}$. The p-value is the probability of being wrong when
denying the normal nature of the fluctuations. Both the SAID
fit~\cite{SAID} and the Granada-All fit provide a similar value which
is outside a $25\sigma$ band, which implies a p-value smaller than
$10^{-10}$. As we discussed in detail in our previous work, one can
tolerate a value of $\chi^2/N_{\rm dat}$ outside the interval $1\pm
\sqrt{2/N_{\rm dat}}$ as long as we can identify a scaled gaussian
distribution by using normality tests.~\cite{Perez:2014kpa}. As it was
pointed out ~\cite{Perez:2014kpa} this is {\it not} the case for the
Granada-All database, and thus the selection of data seems
mandatory~\footnote{ We ignore to what extent our conclusions hold
  also for the SAID analysis, but to our knowledge the normality of
  residuals of the SAID fit has never been reported. Thus, the
  interesting possibility of globally scaling the errors remains to be
  established.}. From the figure it is also clear that the
self-consistent Granada-2013 database is so far the largest database
consistent with a statistically successfull PWA of np and pp
scattering below LAB energy $350 {\rm MeV}$.

\begin{figure*}
\centering
\includegraphics[width=\linewidth]{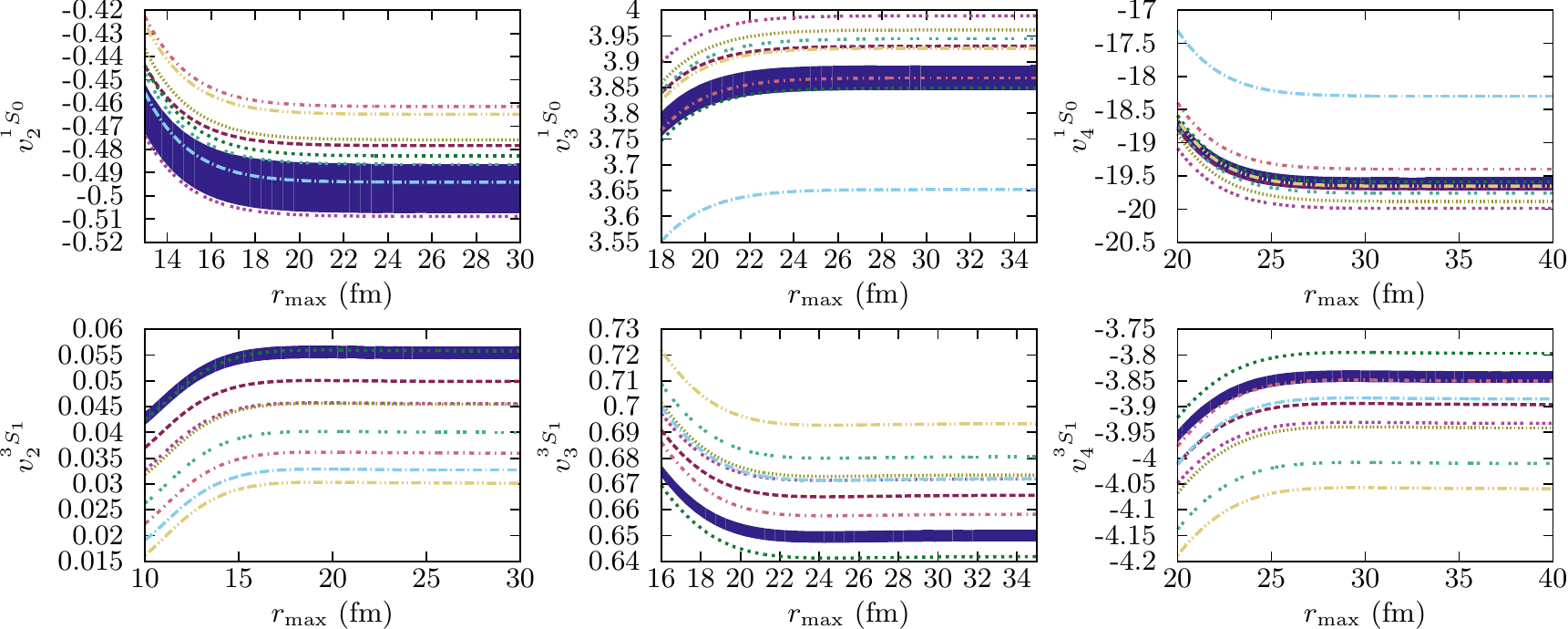}
\caption{Convergence of the $^1S_0$ and $^ 3S_1$ low energy threshold
  parameters $v_2$, $v_3$ and $v_4$ as a function of the integration
  distance and for the DS-OPE~\cite{Perez:2013mwa,Perez:2013jpa} (blue
  band), DS-$\chi$TPE~\cite{Perez:2013oba,Perez:2013cza} (dashed red
  line), Gauss-OPE~\cite{Perez:2014yla} (dotted green line),
  Gauss-$\chi$TPE (dotted light green line), DS-$\Delta$BO (dotted
  purple line), Gauss-$\Delta$BO (dotted dashed ligth yellow line),
  NijmII~\cite{Stoks:1994wp} (dotted olive green line),
  Reid93~\cite{Stoks:1994wp} (dotted dashed light blue line) and
  AV18~\cite{Wiringa:1994wb} (dotted dashed light red line). The width
  of the blue band reflects the corresponding statistical error estimate.}
\label{fig:v2v3v4}       
\end{figure*}

\section{Low energy expansion}
\label{sec:low-exp}

As already mentioned, in the absence of complete sets of measurements
one must resort to specific potentials to carry out the PWA. Since the
form of the potential is chosen and fixed {\it a priori}, the analysis
of NN scattering data is subjected to inverse scattering ambiguities
which are amplified as the energy increases. Therefore one expects
lowest energy information to be more universal and thus we use the
effective range expansion (ERE)~\cite{Bethe:1949yr} as the suitable
tool. Although for S-waves the calculation of the low energy threshold
parameters is straightforward and even customary for NN potentials,
their calculation for higher and coupled channel partial waves is a
computational challenge which has seldomly been addressed.  To
start with, there are not even ready-to-use formulas in the coupled
channel case, and a low energy expansion of the wave function to high
orders is needed. In addition, partial waves with high angular
momentum become numerically unstable as the main contribution comes
from very long distances requiring demanding numerical computations.
This is the reason why these low energy parameters have been very
rarely computed or, when they have been, a very limiting accuracy has
been displayed.  Using Calogero's variable phase approach to the full
S-matrix~\cite{calogero1967variable} the low energy parameters have
only been calculated for the Reid93 and NijmII potentials up to $j \le
5$~\cite{PavonValderrama:2005ku}.  Here we improve on the accuracy of
that work and determine for the first time the statistical
uncertainties of the low energy threshold parameters. 
The method
  to compute these low energy parameters in the general case is based
  on the discrete variable-S-matrix method which we explain in more
  detail in Appendix~\ref{sec:discrete-S}. In essence in this method
  one replaces the original potential by a sum of delta-shells, which
  in the equidistant case corresponds to $U(r) \to \sum_{i=1} U(r_i)
  \delta (r-r_i) \Delta r $ and computes the accumulated phase of
  S-matrix es the number of grid points is switched on. 
 The low
energy parameters for higher partial waves determined in
Ref.~\cite{PavonValderrama:2005ku} have been used to implement
renormalization
conditions~\cite{Arriola:2010hj,Valderrama:2011mv,Long:2011qx,Long:2011xw}
and to analyze causality bounds in np
scattering~\cite{Elhatisari:2012ym}.

In the well known case of central $^1S_0$ and $^3S_1$ partial waves
the ERE is given by (using the nuclear bar representation)
\begin{equation}
 k \cot \delta_0 = - \frac{1}{\alpha_0} + \frac{1}{2}r_0 k^2 + v_2 k^4 +
v_3 k^6 + v_4 k^8 + \ldots,
\end{equation}
where $k$ is the center of mass momentum, $\delta_0$ is the
corresponding partial wave phase-shift, $\alpha_0$ is the scattering
length, $r_0$ is the effective range and $v_i$ are known as the
curvature parameters. The generalization to N-coupled partial waves
with angular momenta $(l_1, \cdots, l_N)$ can be done by introducing
the $\mathbf{\hat{M}}$ matrix defined as
\begin{equation}
 \mathbf{DSD}^{-1} = \left(\mathbf{\hat{M}}+ik\mathbf{D}^2\right)\left(\mathbf{\hat{M}}-ik\mathbf{D}^2\right)^{-1},
\end{equation}
where $\mathbf{S}$ is the usual unitary S-matrix and $\mathbf{D}={\rm
  diag}(k^{l_1},\ldots,k^{l_N})$. In the limit $k \to 0$
the $\mathbf{\hat{M}}$-matrix becomes
\begin{equation}
\label{MLowE}
 \mathbf{\hat{M}} = -\mathbf{a}^{-1} + \frac{1}{2} \mathbf{r} k^2 +
\mathbf{v_2} k^4 + \mathbf{v_3} k^6 + \mathbf{v_4} k^8 + \dots,
\end{equation}
where $\mathbf{a}$, $\mathbf{r}$ and $\mathbf{v_i}$ are the coupled
channel generalizations of $\alpha_0$, $r_0$ and $v_i$ respectively.
Due to $n\pi$ exchange $\mathbf{\hat{M}}(k)$ has branch cuts at $k=\pm
i n m_\pi/2$, and thus the ERE converges for $|k| < m_\pi/2$ or
$E_{\rm LAB} \lesssim 10 {\rm MeV}$. Conversely, the ERE to {\it
  finite} order does not allow to reconstruct the full functions in
the complex plane without the explicit cut structure information.

In the NN case these matrices have dimension 1 and 2. We refer to
Ref.~\cite{PavonValderrama:2005ku} for further details.
In the interesting case of the $^3S_1$ eigen-channel, one
has~\footnote{We use the notation $v=v_2$,$v'=v_3$ and $v''=v_4$ for
simplicity}
\begin{eqnarray}
k \cot \delta_{3S1}^{\rm Eigen} = -\frac1{\alpha_{3S1}^{\rm Eigen}}+ \frac12 r_{3S1}^{\rm Eigen}
k^2 + v_{3S1}^{\rm Eigen} k^4 + \dots
\label{eq:keigen}
\end{eqnarray}  
where we get for the effective range parameters the relations between
the eigen and the nuclear bar (denoted emphatically as barred)
representations,
\begin{eqnarray}
\alpha_{3S1}^{\rm Eigen} &=& \bar \alpha_{3S1}  \\ 
r_{3S1}^{\rm Eigen} &=& \bar
r_{3S1}+ \frac{2 \bar r_{E1} \bar \alpha_{E1}}{\bar \alpha_{3S1}}+
\frac{ \bar r_{3D1} \bar \alpha_{E1}^2}{\bar \alpha_{3S1}^2 } \\
v_{3S1}^{\rm Eigen} &=& \bar v_{3S1} + \frac14 \bar \alpha_{3D1} \bar r_{E1}^2
\nonumber \\ 
&+& \frac{\bar \alpha_{E1}}{4 \bar \alpha_{3S1}} \left (
2 \bar \alpha_{3D1} \bar r_{3D1} \bar r_{E1} - \bar \alpha_{E1} \bar
r_{E1}^2 + 8 \bar v_{E1} \right) \nonumber \\
 &+& \frac{\bar \alpha_{E1}^2}{4\bar \alpha_{3S1}^2} \left( 
\bar \alpha_{3D1} \bar r_{3D1}^2 - 2 
 \bar \alpha_{E1} \bar r_{3D1} \bar r_{E1} + 4 \bar v_{3D1} \right)
 \nonumber \\ 
&+& \frac1{4 \bar \alpha_{3S1}^3} \left( 4 \bar \alpha_{E1}^2 - \bar
 \alpha_{E1}^4 \bar r_{3D1}^2 \right)
\label{eq:sym-eigen} 
\end{eqnarray} 
and so on.

\begin{table}[htb]
\caption{\small Low-energy scattering parameters for the $^{3}S_{1}$
  eigen-phase. We quote the numbers of Ref.~\cite{deSwart:1995ui} for
  the PWA~\cite{Stoks:1993tb} and the Nijm-I, Nijm-II and Reid 93
  potentials~\cite{Stoks:1994wp}, (first four rows), our results
  integrating the discrete variable S-matrix equations with $N=2
  \times 10^5$ grid points for the Nijm-II and Reid
  93~\cite{Stoks:1994wp} and AV18~\cite{Wiringa:1994wb} potentials
  quoting numerical errors (in boldface) relative to 
the computation with $N=1
 \times 10^5$. Statistical errors are 
also quoted when available.} \label{tab:Nijm}
{\footnotesize
\begin{center}
\renewcommand{\arraystretch}{1.2}
\begin{tabular}{l|llllll}
  &\multicolumn{1}{c}{$\alpha_0$}&\multicolumn{1}{c}{$r_0$}
  &\multicolumn{1}{c}{$v_{2}$}&\multicolumn{1}{c}{$v_{3}$}
  &\multicolumn{1}{c}{$v_{4}$}  \\\hline
PWA     & 5.420(1) & 1.753(2) & 0.040 & 0.672 &--3.96  \\[1mm]
Nijm~I  & 5.418    & 1.751    & 0.046 & 0.675 &--3.97  \\
Nijm~II  & 5.420    & 1.753    & 0.045 & 0.673 &--3.95  \\
Reid93   & 5.422    & 1.755    & 0.033 & 0.671 &--3.90  \\
\hline 
NijmII &5.4197({\bf 3})&1.75343({\bf 3})&0.04545({\bf 1})& 0.6735({\bf 1})& --3.9414({\bf 8}) \\
Reid93 &5.4224({\bf 2})&1.75550({\bf 3})&0.03269({\bf 1})&0.6721({\bf 1})& --3.8867({\bf 7}) \\ 
AV18   &5.4020({\bf 2})&1.75171({\bf 3})&0.03598({\bf 1})&0.6583({\bf 1})& --3.8507({\bf 7}) \\
\hline 
DS-OPE         & 5.435(2)    & 1.774(3)    & 0.055(1) & 0.650(2) &--3.84(1)  \\ 
DS-TPE         & 5.424       & 1.760       & 0.050    & 0.666    &--3.90 \\ 
DS-$\Delta$BO  & 5.419       & 1.752       & 0.046    & 0.672    &--3.93 \\
G-OPE          & 5.441       & 1.781       & 0.056    & 0.642    &--3.80 \\ 
G-TPE          & 5.410       & 1.739       & 0.040    & 0.681    &--4.01 \\
G-$\Delta$BO   & 5.397       & 1.722       & 0.030    & 0.693    &--4.06 
\end{tabular}
\end{center}}
\end{table}

\begin{table}
 \caption{\label{tab:LEPS-statistic} Low energy threshold np
   parameters for all partial waves with $j \leq 5$. The central value
   and \emph{statistical} error bars are given on the first line of
   each partial wave and correspond to the mean and standard deviation
   of a population of $1020$ parameters calculated with the Monte
   Carlo family of potential parameters described
   in~\cite{Perez:2014jsa} using the DS-OPE
   potential~\cite{Perez:2013mwa,Perez:2013jpa}.  The second line
   quotes the \emph{systematic} uncertainties, the central value and
   error bars correspond to the mean and standard deviation of the 9
   realistic potentials NijmII~\cite{Stoks:1994wp},
   Reid93~\cite{Stoks:1994wp}, AV18~\cite{Wiringa:1994wb},
   DS-OPE~\cite{Perez:2013mwa,Perez:2013jpa},
   DS-$\chi$TPE~\cite{Perez:2013oba,Perez:2013cza},
   Gauss-OPE~\cite{Perez:2014yla}, Gauss-$\chi$TPE, DS-$\Delta$BO and
   Gauss-$\Delta$BO. For each partial wave we show the scattering
   length $\alpha$ and the effective range $r_0$, both in ${\rm
     fm}^{l+l'+1}$, as well as the curvature parameters $v_2$ in ${\rm
     fm}^{l+l'+3}$, $v_3$ in ${\rm fm}^{l+l'+5}$ and $v_4$ in ${\rm
     fm}^{l+l'+5}$. For the coupled channels we use the nuclear bar
   representation of the $S$ matrix. Uncertainties smaller than
   $10^{-3}$ are not quoted} {\footnotesize
  \begin{tabular*}{\columnwidth}{@{\extracolsep{\fill}} l D{.}{.}{3.5}
 D{.}{.}{3.5} D{.}{.}{3.5} D{.}{.}{3.5} D{.}{.}{4.7} }
 Wave   & \multicolumn{1}{c}{$\alpha$} 
        & \multicolumn{1}{c}{$ r_0$} 
        & \multicolumn{1}{c}{$ v_2$} 
        & \multicolumn{1}{c}{$ v_3$} 
        & \multicolumn{1}{c}{$ v_4$} \\
 \hline \noalign{\smallskip}
$^1S_0$ &   -23.735(6)  &    2.673(9)  &    -0.50(1)  &     3.87(2)  &   -19.6(1)   \\
        &   -23.735(16) &    2.68(3)   &    -0.48(2)  &     3.9(1)   &   -19.6(5)   \\
$^3P_0$ &    -2.531(6)  &    3.71(2)   &     0.93(1)  &     3.99(3)  &    -8.11(5)  \\
        &    -2.5(1)    &    3.7(4)    &     0.9(5)   &     3.9(1)   &    -8.2(9)   \\
$^1P_1$ &     2.759(6)  &   -6.54(2)   &    -1.84(5)  &     0.41(2)  &     8.39(9)  \\
        &     2.78(3)   &   -6.46(9)   &    -1.7(2)   &     0.5(2)   &     8.0(3)   \\
$^3P_1$ &     1.536(1)  &   -8.50(1)   &     0.02(1)  &    -1.05(2)  &     0.56(1)  \\
        &     1.52(1)   &   -8.6(1)    &    -0.06(7)  &    -0.9(2)   &     0.1(5)   \\
$^3S_1$ &     5.435(2)  &    1.852(2)  &    -0.122(3) &     1.429(7) &    -7.60(3)  \\
        &     5.42(1)   &    1.84(1)   &    -0.14(1)  &     1.46(3)  &    -7.7(2)   \\
$\EP_1$ &     1.630(6)  &    0.400(3)  &    -0.266(5) &     1.47(1)  &    -7.28(2)  \\
        &     1.61(2)   &    0.39(2)   &    -0.29(3)  &     1.47(2)  &    -7.35(9)  \\
$^3D_1$ &     6.46(1)   &   -3.540(8)  &    -3.70(2)  &     1.14(2)  &    -2.77(2)  \\
        &     6.43(4)   &   -3.57(2)   &    -3.77(4)  &     1.11(5)  &    -2.7(1)   \\
$^1D_2$ &    -1.376     &   15.04(2)   &    16.68(6)  &   -13.5(1)   &    35.4(1)   \\
        &    -1.379(6)  &   15.00(9)   &    16.7(2)   &   -12.9(4)   &    36.2(14)  \\
$^3D_2$ &    -7.400(4)  &    2.858(3)  &     2.382(9) &    -1.04(2)  &     1.74(2)  \\
        &    -7.39(1)   &    2.87(1)   &     2.41(3)  &    -0.96(5)  &     1.75(8)  \\
$^3P_2$ &    -0.290(2)  &   -8.19(1)   &    -6.57(5)  &    -5.5(2)   &   -12.2(3)   \\
        &    -0.288(5)  &   -8.3(2)    &    -6.8(7)   &    -6.1(19)  &   -12.7(26)  \\
$\EP_2$ &     1.609(1)  &  -15.68(2)   &   -24.91(8)  &   -21.9(3)   &   -64.1(7)   \\
        &     1.604(6)  &  -15.8(2)    &   -25.2(7)   &   -23.0(29)  &   -66.2(69)  \\
$^3F_2$ &    -0.971     &   -5.74(2)   &   -23.26(8)  &   -79.5(4)   &  -113.0(16)  \\
        &    -0.971(5)  &   -5.7(1)    &   -23.3(6)   &   -80.1(33)  &  -117.2(121) \\
$^1F_3$ &     8.378     &   -3.924     &    -9.869(4) &   -15.27(2)  &    -1.95(7)  \\
        &     8.377(5)  &   -3.926(4)  &    -9.88(2)  &   -15.3(1)   &    -2.2(4)   \\
$^3F_3$ &     2.689     &   -9.978(3)  &   -20.67(2)  &   -19.12(8)  &   -27.7(2)   \\
        &     2.690(6)  &   -9.97(2)   &   -20.65(8)  &   -19.0(3)   &   -26.9(7)   \\
$^3D_3$ &    -0.134     &    1.373     &     2.082(3) &     1.96(1)  &    -0.45(3)  \\
        &    -0.14(2)   &    1.371(3)  &     2.07(1)  &     1.92(6)  &    -0.51(9)  \\
$\EP_3$ &    -9.682     &    3.262     &     7.681(3) &     9.62(2)  &    -1.09(5)  \\
        &    -9.683(5)  &    3.260(5)  &     7.67(2)  &     9.6(1)   &    -1.1(2)   \\
$^3G_3$ &     4.876     &   -0.027     &     0.019(2) &     0.07(1)  &    -2.69(3)  \\
        &     4.875(3 ) &   -0.03(1)   &    -0.01(6)  &    -0.05(30) &    -2.8(7)   \\
$^1G_4$ &    -3.208     &   10.833(1)  &    34.629(9) &    83.04(8)  &   108.1(4)   \\
        &    -3.212(6)  &   10.81(2)   &    34.53(7)  &    82.4(4)   &   105.6(15)  \\
$^3G_4$ &   -19.145     &    2.058     &     6.814    &    16.769(4) &    10.00(2)  \\
        &   -19.147(8)  &    2.058     &     6.815(3) &    16.78(2)  &    10.04(6)  \\
$^3F_4$ &    -0.006     &   -3.043     &    -4.757(1) &    73.903(5) &   662.21(9)  \\
        &    -0.010(2)  &   -3.044(8)  &    -4.77(5)  &    73.9(3)   &   662.8(32)  \\
$\EP_4$ &     3.586     &   -9.529     &   -37.02(3)  &  -184.40(2)  &  -587.28(9)  \\
        &     3.589(8)  &   -9.53(2)   &   -37.04(7)  &  -184.6(3)   &  -586.6(17)  \\
$^3H_4$ &    -1.240     &   -0.157(2)  &    -1.42(1)  &   -14.0(1)   &   -99.0(9)   \\
        &    -1.241(3)  &   -0.18(1)   &    -1.55(9)  &   -15.2(8)   &  -106.7(55)  \\
$^1H_5$ &    28.574     &   -1.727     &    -7.906    &   -32.787    &   -59.361    \\
        &    28.58(1)   &   -1.727     &    -7.906(4) &   -32.78(2)  &   -59.38(5)  \\
$^3H_5$ &     6.081     &   -6.439     &   -25.228    &   -82.511(3) &  -168.47(2)  \\
        &     6.09(2)   &   -6.44(1)   &   -25.22(5)  &   -82.5(1)   &  -168.3(8)   \\
$^3G_5$ &    -0.008     &    0.481     &     1.878    &     6.100    &     6.791    \\
        &    -0.009(1)  &    0.480     &     1.878    &     6.098(3) &     6.784(9) \\
$\EP_5$ &   -31.302     &    1.556     &     6.995    &    28.179    &    48.376(2) \\
        &   -31.31(1)   &    1.556     &     6.993(3) &    28.17(1)  &    48.35(3)  \\
$^3I_5$ &    10.678     &    0.011     &     0.146    &     1.441    &     6.546(6) \\
        &    10.680(5)  &    0.011     &     0.145(1) &     1.43(1)  &     6.47(9)   \\
 \end{tabular*}
}
\end{table}

\section{Sampling the NN interaction: Fine graining vs Coarse graining}
\label{sec:fine-coarse}

Several high-quality interactions stemming from the 1993 Nijmegen PWA
such as the NijmII, Reid93, AV18 potentials are smooth functions in
configuration space. For the discrete S-matrix method (see
  Appendix~\ref{sec:discrete-S} for details) this means that the
values $U(r_i)$ are {\it given} for $U=U_{\rm Nijm II}, U_{\rm
  Reid93}, U_{\rm AV18}$, and thus a fine graining $\Delta r \to 0$ is
needed. We test the numerical accuracy and precision of the approach
by using a finite grid representation and determine the low energy
parameters of these potentials. In particular, we take $r_{\rm
  max}=100 \, {\rm fm}$ and $ \Delta r = 0.01, 0.005,0.001,0.0005 \,
{\rm fm}$ corresponding to $N= 1 \times 10^4, 2 \times 10^4, 10^5, 2
\times 10^5$ grid points respectively and convergence is established
for both $\Delta r$ and $r_{\rm max}$.

For illustration we show in Fig.~\ref{fig:v2v3v4} eye-ball convergence
for $v_{2,3,4}$ (eigen) in the  $^3S_1$ channel, which is achieved with an
integration upper limit of $r_{\rm max}=30-40 {\rm fm}$.  Sufficiently
high numerical convergence is comfortably obtained with $r_{\rm max}=100
{\rm fm}$ for all partial waves with $J \le 5$.

In order to gauge the accuracy of our calculation we compare with the
last revision of the Nijmegen
group~\cite{deSwart:1995ui}. In table~\ref{tab:Nijm} we show our
results computed with the DVSM method. As we see, our implementation
allows for high numerical precision which can be tuned to be the
highest one among other sources of uncertainties, namely statistical
and systematic errors to be discussed below.  If we take the quoted
numbers in Ref.~\cite{deSwart:1995ui} as significant figures, and
assuming the standard round-off error rules, their numerical error is
smaller than a half of the last provided digit. Our results are mostly
compatible with theirs but considerably more precise.

It is useful to ponder on our numerical accuracy by looking into other
possible integration methods.  In the conventional Numerov of
Runge-Kutta methods, usually employed for smooth
potentials, convergence is defined in terms of the precision of the
wave function, so that one needs a large number of mesh-points. 
The accuracy is also an issue in momentum space
calculations where the momentum grid has an ultraviolet cut-off $\Delta p$
which requires large matrices to make the low energy limit precise.

Alternatively, as pointed out in our previous works, the {\it same}
level of accuracy and precision with much less computational cost can
be achieved by taking the $U(r_i)$ as fitting parameters themselves to
NN scattering data (or even phase-shifts or scattering
amplitudes). This is the basic idea behind coarse graining, implicit
in the work by Avil\'es~\cite{Aviles:1973ee} and exploited in
Refs.~\cite{Perez:2013mwa,Perez:2013jpa} as the DS-potential samples
the interaction with an integration step fixed by the maximum
resolution dictated by the shortest de Broglie wavelength, namely
$\Delta r \sim 0.5$ fm.  A further advantage as compared to more
conventional methods is the numerical stability of the method, since
the number of arithmetic operations required with a few delta shells
avoids accumulation of round-off errors.

\section{Statistical and Systematic Uncertainties}
\label{sec:stat-sys}

The uncertainty discussed in the previous section for some low energy
parameters is purely of numerical character and does not reflect the
physical accuracy inferred directly from the experimental
data~\cite{Perez:2013mwa, Perez:2013jpa} nor the dependence inherited
from the model used to analyze the data. In what follows we analyze
the statistical and systematic uncertainties for the low energy
parameters, the scattering phase shifts and the 5 complex scattering
amplitudes in terms of the Wolfenstein parameters.

Statistical uncertainties are presented in
table~\ref{tab:LEPS-statistic} which shows the low energy np threshold
parameters of all partial waves with $j \leq 5$ for the DS-OPE
potential presented in~\cite{Perez:2013mwa, Perez:2013jpa}. To
propagate statistical uncertainties we use our recent Monte Carlo
bootstrap to NN data~\cite{Perez:2014jsa}, where the set of potential
parameters is replicated $1020$ times, and the mean and standard
deviation provide the central value and $1\sigma$ confidence interval
respectively. It is very important to note that even though the
threshold parameters encode the low energy structure of the NN
interaction the statistical uncertainties are propagated from
scattering data up to $350$MeV. This approach encodes the high
accuracy of a full-fledged PWA into a model independent low energy
representation featured by the ERE. As we see the statistical
precision is very high. We note in passing that, compared to our
analysis of the $^3S_1$-eigen channel, the Nijmegen group had about
$70\%$ the data but provided {\it twice} the statistical precision as
we do (see Table~\ref{tab:Nijm}). This apparent inconsistency could be
due to the different error propagation method.

\begin{figure*}
\centering
\includegraphics[width=\linewidth]{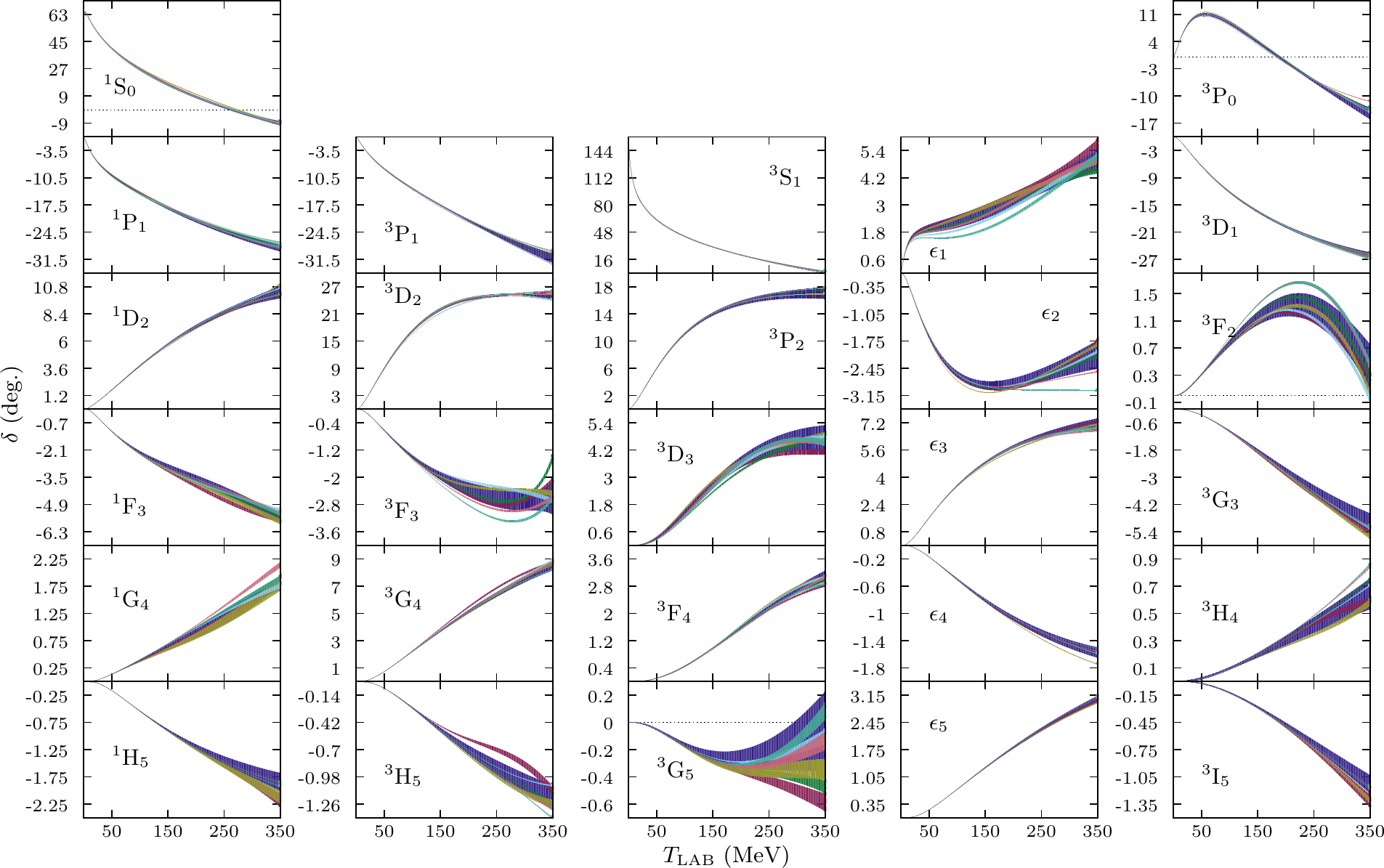}
\caption{np phase-shifts in degrees for all partial waves with $J \leq
  5$. The dark blue band represents the mean and standard deviation of
  thirteen different determinations of the NN interaction to their
  contemporary database~\cite{Stoks:1993tb, Stoks:1994wp,
    Wiringa:1994wb, Machleidt:2000ge, Gross:2008ps, Perez:2013mwa,
    Perez:2013jpa, Perez:2013oba, Perez:2013cza, Perez:2014yla}. The
  red, green, olive green, light blue, light red and light green bands
  represent the statistical uncertainty of the
  DS-OPE~\cite{Perez:2013mwa,Perez:2013jpa},
  DS-$\chi$TPE~\cite{Perez:2013oba,Perez:2013cza},
  Gauss-OPE~\cite{Perez:2014yla}, Gauss-$\chi$TPE, DS-$\Delta$BO and
  Gauss-$\Delta$BO potentials respectively.}
\label{fig:PSSystematic}       
\end{figure*}

\begin{figure*}
\centering
\includegraphics[width=\linewidth]{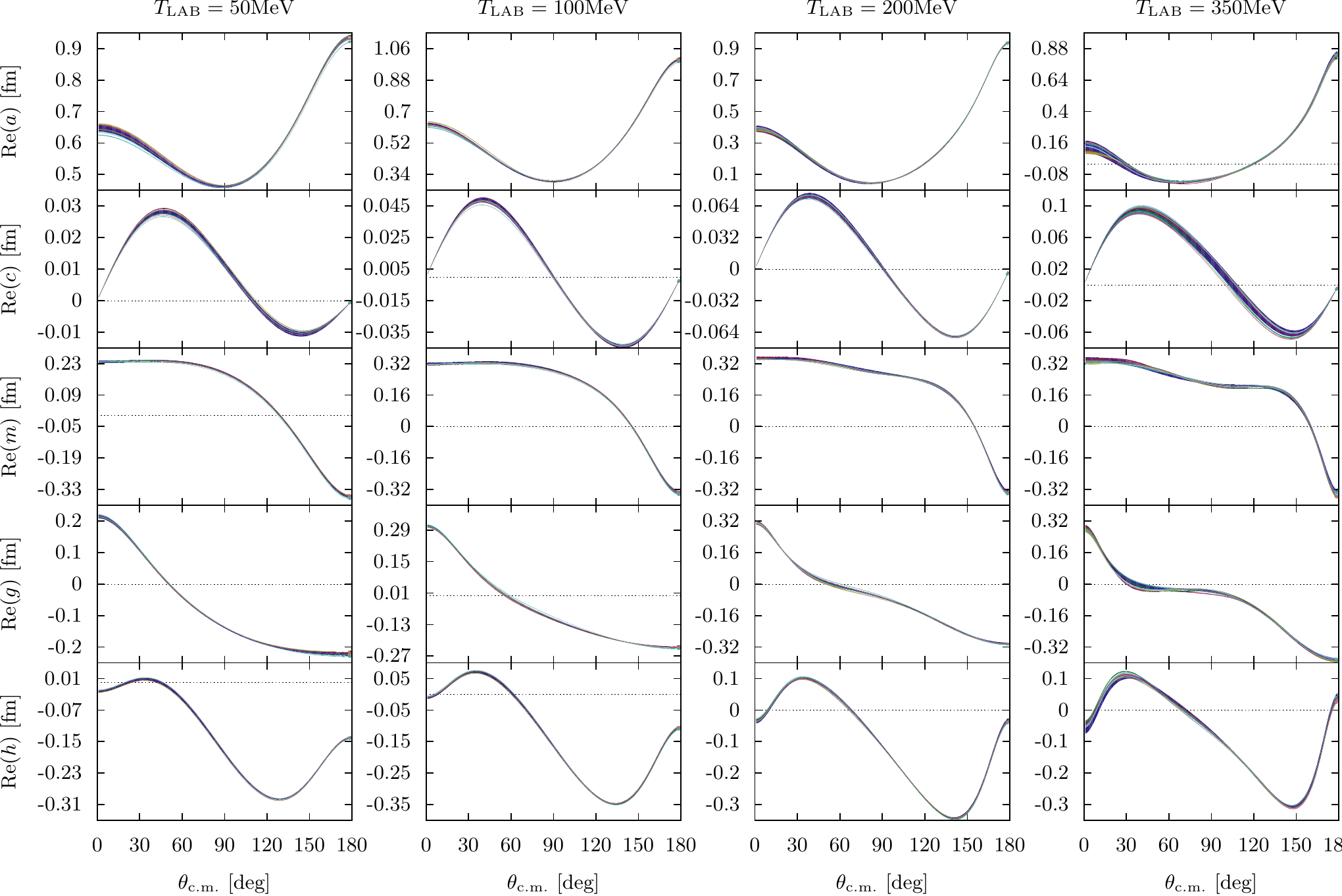}
\caption{Same as Fig.~\ref{fig:PSSystematic} for the real part of the
  Wolfenstein parametrization of the np scattering amplitude in fm as
  a function of center of mass scattering angle at $T_{\rm LAB} = 50,
  \ 100, \ 200, \ 350$MeV.}
\label{fig:ReWolfenstein}       
\end{figure*}

\begin{figure*}
\centering
\includegraphics[width=\linewidth]{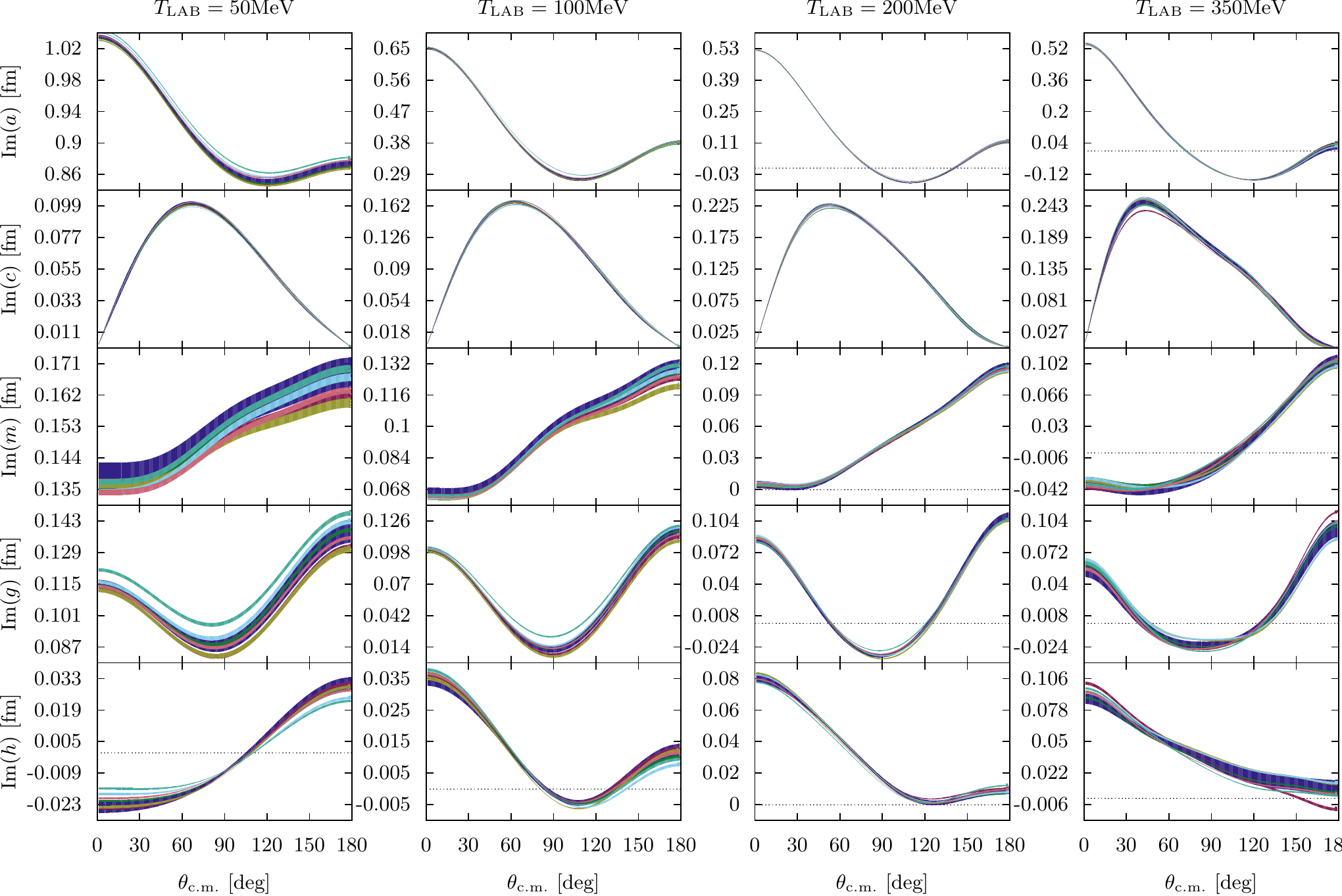}
\caption{Same as Fig.~\ref{fig:ReWolfenstein} for the imaginary part.}
\label{fig:ImWolfenstein}       
\end{figure*}

We turn now to estimate the systematic uncertainty. Even though
several phenomenological potentials can reproduce their contemporary
NN scattering database, discrepancies have been found when comparing
their corresponding phase-shifts~\cite{NavarroPerez:2012vr,
  Perez:2012kt}. In Fig.~\ref{fig:PSSystematic} we show the np
phase-shifts up to $T_{\rm LAB} = 350$MeV for all partial waves with
$J \leq 5$. The systematic uncertainty is represented as a band
indicating the mean and standard deviation of thirteen high-quality
determinations of the np interaction, in particular the PWA from the
Nijmegen group~\cite{Stoks:1993tb}, the NijmI, NijmII,
Reid93~\cite{Stoks:1994wp}, AV18~\cite{Wiringa:1994wb} and
CD-Bonn~\cite{Machleidt:2000ge} potentials, the covariant spectator
model~\cite{Gross:2008ps} and our model potentials
DS-OPE~\cite{Perez:2013mwa,Perez:2013jpa},
DS-$\chi$TPE~\cite{Perez:2013oba,Perez:2013cza},
Gauss-OPE~\cite{Perez:2014yla}, Gauss-$\chi$TPE, DS-$\Delta$BO and
Gauss-$\Delta$BO.  We note larger discrepancies in the mixing
  $\epsilon_1$ parameter corresponding to the $^3S_1-^3D_1$ (deuteron)
  channel as well as the peripheral $^3F_2$ and $^3G_5$ waves (note,
  however, their smallness in comparison to other partial
    waves).  These thirteen determinations were not made with the
same database and therefore could not be used collectively to
determine the systematic uncertainty. However, the DS-OPE,
DS-$\chi$TPE, DS-$\Delta$BO, Gauss-OPE, Gauss-$\chi$TPE and
Gauss-$\Delta$BO potentials are fitted to the same self-consistent
database with normally distributted residuals, but their phase-shifts
with statitiscal uncertainties, shown also in
Fig.~\ref{fig:PSSystematic}, do not always overlap.  Their
discrepancies are of the same order of the systematic uncertainty
band. The additional data, while reducing the statistical uncertainty,
do not modify the systematic uncertainty. A thorough study of the
data distribution on the $(T_{\rm LAB},\theta_{\rm c.m.})$-plane could
provide meaningfull information on which scattering measurements are
necessary to reduce the systematic uncerntainties by avoiding an
abundance bias. Although the propagation of systematic uncertainties
is not as direct as the statistical one, we observe that differences
in phase-shifts tend to be at least an order of magnitude larger than
the statistical error bars~\cite{Perez:2013oba, Perez:2013cza,
  Perez:2014yla}. A similar trend is found when comparing scattering
amplitudes as can be seen in Fig.~\ref{fig:ReWolfenstein} and
Fig.~\ref{fig:ImWolfenstein}.  Again the width of blue band
  representing the systematic uncertainty is always about an order of
  magnitude larger than the statistical one. This is more evident in
  Fig.~\ref{fig:ImWolfenstein} where the scale on the $y$-axis allows
  for a clear comparison of all bands.

\begin{table*}
 \footnotesize
 \caption{\label{tab:ppIsovectorPS} Averaged pp isovector phaseshifts in degrees (errors are systematic).}
 \begin{ruledtabular}
 \begin{tabular*}{\textwidth}{@{\extracolsep{\fill}} r *{12}{D{.}{.}{3.3}}}
$E_{\rm LAB}$&\multicolumn{1}{c}{$^1S_0$}&\multicolumn{1}{c}{$^1D_2$}&\multicolumn{1}{c}{$^1G_4$}&\multicolumn{1}{c}{$^3P_0$}&\multicolumn{1}{c}{$^3P_1$}&\multicolumn{1}{c}{$^3F_3$}&\multicolumn{1}{c}{$^3P_2$}&\multicolumn{1}{c}{$\epsilon_2$}&\multicolumn{1}{c}{$^3F_2$}&\multicolumn{1}{c}{$^3F_4$}&\multicolumn{1}{c}{$\epsilon_4$}&\multicolumn{1}{c}{$^3H_4$}\\ 
  \hline 
  1 &    32.677 &     0.001 &     0.000 &     0.133 &    -0.079 &    -0.000 &     0.014 &    -0.001 &     0.000 &     0.000 &    -0.000 &     0.000\\
    & \pm 0.016 & \pm 0.000 & \pm 0.000 & \pm 0.000 & \pm 0.001 & \pm 0.000 & \pm 0.000 & \pm 0.000 & \pm 0.000 & \pm 0.000 & \pm 0.000 & \pm 0.000\\
  5 &    54.895 &     0.042 &     0.000 &     1.575 &    -0.888 &    -0.004 &     0.214 &    -0.052 &     0.002 &     0.000 &    -0.000 &     0.000\\
    & \pm 0.037 & \pm 0.000 & \pm 0.000 & \pm 0.006 & \pm 0.006 & \pm 0.000 & \pm 0.005 & \pm 0.000 & \pm 0.000 & \pm 0.000 & \pm 0.000 & \pm 0.000\\
 10 &    55.320 &     0.164 &     0.003 &     3.717 &    -2.028 &    -0.031 &     0.648 &    -0.200 &     0.013 &     0.001 &    -0.003 &     0.000\\
    & \pm 0.065 & \pm 0.000 & \pm 0.000 & \pm 0.017 & \pm 0.015 & \pm 0.000 & \pm 0.012 & \pm 0.001 & \pm 0.000 & \pm 0.000 & \pm 0.000 & \pm 0.000\\
 25 &    48.848 &     0.689 &     0.040 &     8.552 &    -4.840 &    -0.231 &     2.479 &    -0.808 &     0.106 &     0.020 &    -0.049 &     0.004\\
    & \pm 0.106 & \pm 0.003 & \pm 0.000 & \pm 0.067 & \pm 0.034 & \pm 0.001 & \pm 0.028 & \pm 0.004 & \pm 0.001 & \pm 0.001 & \pm 0.000 & \pm 0.000\\
 50 &    39.182 &     1.685 &     0.153 &    11.436 &    -8.161 &    -0.692 &     5.837 &    -1.707 &     0.345 &     0.108 &    -0.197 &     0.026\\
    & \pm 0.115 & \pm 0.017 & \pm 0.001 & \pm 0.169 & \pm 0.044 & \pm 0.008 & \pm 0.039 & \pm 0.015 & \pm 0.010 & \pm 0.006 & \pm 0.000 & \pm 0.000\\
100 &    25.357 &     3.750 &     0.423 &     9.324 &   -13.109 &    -1.530 &    11.027 &    -2.675 &     0.853 &     0.471 &    -0.549 &     0.111\\
    & \pm 0.136 & \pm 0.051 & \pm 0.006 & \pm 0.327 & \pm 0.085 & \pm 0.048 & \pm 0.052 & \pm 0.036 & \pm 0.054 & \pm 0.016 & \pm 0.003 & \pm 0.002\\
150 &    15.229 &     5.639 &     0.706 &     4.532 &   -17.379 &    -2.142 &    14.059 &    -2.954 &     1.271 &     1.011 &    -0.868 &     0.221\\
    & \pm 0.231 & \pm 0.047 & \pm 0.004 & \pm 0.383 & \pm 0.135 & \pm 0.135 & \pm 0.062 & \pm 0.032 & \pm 0.117 & \pm 0.020 & \pm 0.008 & \pm 0.011\\
200 &     7.076 &     7.212 &     1.005 &    -0.494 &   -21.273 &    -2.568 &    15.768 &    -2.914 &     1.499 &     1.628 &    -1.132 &     0.343\\
    & \pm 0.281 & \pm 0.087 & \pm 0.014 & \pm 0.313 & \pm 0.252 & \pm 0.265 & \pm 0.118 & \pm 0.041 & \pm 0.183 & \pm 0.037 & \pm 0.012 & \pm 0.028\\
250 &     0.212 &     8.487 &     1.311 &    -5.160 &   -24.784 &    -2.809 &    16.721 &    -2.724 &     1.457 &     2.220 &    -1.340 &     0.474\\
    & \pm 0.264 & \pm 0.107 & \pm 0.041 & \pm 0.147 & \pm 0.527 & \pm 0.377 & \pm 0.203 & \pm 0.131 & \pm 0.230 & \pm 0.054 & \pm 0.026 & \pm 0.053\\
300 &    -5.694 &     9.525 &     1.605 &    -9.287 &   -27.880 &    -2.814 &    17.208 &    -2.460 &     1.093 &     2.710 &    -1.501 &     0.615\\
    & \pm 0.354 & \pm 0.088 & \pm 0.068 & \pm 0.452 & \pm 0.962 & \pm 0.375 & \pm 0.323 & \pm 0.264 & \pm 0.236 & \pm 0.063 & \pm 0.046 & \pm 0.083\\
350 &   -10.828 &    10.375 &     1.865 &   -12.813 &   -30.527 &    -2.447 &    17.376 &    -2.157 &     0.389 &     3.048 &    -1.623 &     0.767\\
    & \pm 0.747 & \pm 0.347 & \pm 0.084 & \pm 1.091 & \pm 1.565 & \pm 0.412 & \pm 0.553 & \pm 0.446 & \pm 0.233 & \pm 0.079 & \pm 0.063 & \pm 0.120\\
 \end{tabular*}
 \end{ruledtabular}
\end{table*}

\begin{table*}
 \footnotesize
 \caption{\label{tab:npIsovectorPS} Averaged np isovector phaseshifts in degrees
   (errors are systematic).}
 \begin{ruledtabular}
  \begin{tabular*}{\textwidth}{@{\extracolsep{\fill}} r *{12}{D{.}{.}{3.3}}}
$E_{\rm LAB}$&\multicolumn{1}{c}{$^1S_0$}&\multicolumn{1}{c}{$^1D_2$}&\multicolumn{1}{c}{$^1G_4$}&\multicolumn{1}{c}{$^3P_0$}&\multicolumn{1}{c}{$^3P_1$}&\multicolumn{1}{c}{$^3F_3$}&\multicolumn{1}{c}{$^3P_2$}&\multicolumn{1}{c}{$\epsilon_2$}&\multicolumn{1}{c}{$^3F_2$}&\multicolumn{1}{c}{$^3F_4$}&\multicolumn{1}{c}{$\epsilon_4$}&\multicolumn{1}{c}{$^3H_4$}\\ 
  \hline 
  1 &    62.105 &     0.001 &     0.000 &     0.178 &    -0.106 &    -0.000 &     0.022 &    -0.001 &     0.000 &     0.000 &    -0.000 &     0.000\\
    & \pm 0.039 & \pm 0.000 & \pm 0.000 & \pm 0.002 & \pm 0.001 & \pm 0.000 & \pm 0.000 & \pm 0.000 & \pm 0.000 & \pm 0.000 & \pm 0.000 & \pm 0.000\\
  5 &    63.689 &     0.041 &     0.000 &     1.626 &    -0.923 &    -0.004 &     0.255 &    -0.048 &     0.002 &     0.000 &    -0.000 &     0.000\\
    & \pm 0.079 & \pm 0.000 & \pm 0.000 & \pm 0.024 & \pm 0.011 & \pm 0.000 & \pm 0.004 & \pm 0.000 & \pm 0.000 & \pm 0.000 & \pm 0.000 & \pm 0.000\\
 10 &    60.038 &     0.155 &     0.002 &     3.672 &    -2.032 &    -0.026 &     0.718 &    -0.183 &     0.011 &     0.001 &    -0.003 &     0.000\\
    & \pm 0.114 & \pm 0.001 & \pm 0.000 & \pm 0.064 & \pm 0.027 & \pm 0.000 & \pm 0.011 & \pm 0.001 & \pm 0.000 & \pm 0.000 & \pm 0.000 & \pm 0.000\\
 25 &    51.011 &     0.670 &     0.032 &     8.250 &    -4.801 &    -0.199 &     2.595 &    -0.753 &     0.091 &     0.017 &    -0.039 &     0.003\\
    & \pm 0.189 & \pm 0.002 & \pm 0.000 & \pm 0.201 & \pm 0.069 & \pm 0.001 & \pm 0.029 & \pm 0.009 & \pm 0.001 & \pm 0.001 & \pm 0.000 & \pm 0.000\\
 50 &    40.644 &     1.686 &     0.134 &    10.955 &    -8.151 &    -0.620 &     5.970 &    -1.635 &     0.310 &     0.098 &    -0.169 &     0.021\\
    & \pm 0.324 & \pm 0.013 & \pm 0.002 & \pm 0.361 & \pm 0.114 & \pm 0.008 & \pm 0.056 & \pm 0.033 & \pm 0.010 & \pm 0.006 & \pm 0.001 & \pm 0.000\\
100 &    26.772 &     3.797 &     0.390 &     8.752 &   -13.218 &    -1.421 &    11.118 &    -2.633 &     0.795 &     0.447 &    -0.499 &     0.094\\
    & \pm 0.620 & \pm 0.044 & \pm 0.018 & \pm 0.479 & \pm 0.184 & \pm 0.046 & \pm 0.086 & \pm 0.071 & \pm 0.051 & \pm 0.013 & \pm 0.005 & \pm 0.002\\
150 &    16.791 &     5.699 &     0.660 &     3.937 &   -17.582 &    -2.028 &    14.086 &    -2.950 &     1.198 &     0.978 &    -0.812 &     0.195\\
    & \pm 0.770 & \pm 0.078 & \pm 0.047 & \pm 0.475 & \pm 0.239 & \pm 0.129 & \pm 0.085 & \pm 0.065 & \pm 0.112 & \pm 0.019 & \pm 0.014 & \pm 0.011\\
200 &     8.759 &     7.262 &     0.945 &    -1.095 &   -21.541 &    -2.462 &    15.736 &    -2.937 &     1.411 &     1.586 &    -1.079 &     0.310\\
    & \pm 0.736 & \pm 0.125 & \pm 0.079 & \pm 0.359 & \pm 0.319 & \pm 0.257 & \pm 0.114 & \pm 0.045 & \pm 0.180 & \pm 0.046 & \pm 0.025 & \pm 0.027\\
250 &     1.982 &     8.516 &     1.248 &    -5.757 &   -25.099 &    -2.713 &    16.638 &    -2.764 &     1.350 &     2.170 &    -1.295 &     0.436\\
    & \pm 0.561 & \pm 0.126 & \pm 0.105 & \pm 0.135 & \pm 0.547 & \pm 0.359 & \pm 0.190 & \pm 0.134 & \pm 0.230 & \pm 0.077 & \pm 0.040 & \pm 0.051\\
300 &    -3.855 &     9.529 &     1.564 &    -9.874 &   -28.229 &    -2.716 &    17.082 &    -2.511 &     0.965 &     2.649 &    -1.465 &     0.573\\
    & \pm 0.357 & \pm 0.128 & \pm 0.121 & \pm 0.430 & \pm 0.958 & \pm 0.329 & \pm 0.329 & \pm 0.283 & \pm 0.231 & \pm 0.097 & \pm 0.059 & \pm 0.081\\
350 &    -8.923 &    10.352 &     1.880 &   -13.387 &   -30.902 &    -2.307 &    17.213 &    -2.217 &     0.239 &     2.975 &    -1.597 &     0.721\\
    & \pm 0.533 & \pm 0.416 & \pm 0.140 & \pm 1.081 & \pm 1.547 & \pm 0.478 & \pm 0.587 & \pm 0.481 & \pm 0.208 & \pm 0.102 & \pm 0.079 & \pm 0.120\\
 \end{tabular*}
 \end{ruledtabular}
\end{table*}

\begin{table*}
 \footnotesize
 \caption{\label{tab:npIsoscalarPS} Averaged np isoscalar phaseshifts in degrees (errors are systematic).}
 \begin{ruledtabular}
 \begin{tabular*}{\textwidth}{@{\extracolsep{\fill}} r *{12}{D{.}{.}{3.3}}}
 $E_{\rm LAB}$&\multicolumn{1}{c}{$^1P_1$}&\multicolumn{1}{c}{$^1F_3$}&\multicolumn{1}{c}{$^3D_2$}&\multicolumn{1}{c}{$^3G_4$}&\multicolumn{1}{c}{$^3S_1$}&\multicolumn{1}{c}{$\epsilon_1$}&\multicolumn{1}{c}{$^3D_1$}&\multicolumn{1}{c}{$^3D_3$}&\multicolumn{1}{c}{$\epsilon_3$}&\multicolumn{1}{c}{$^3G_3$}\\ 
  \hline 
  1 &    -0.188 &    -0.000 &     0.006 &     0.000 &   147.748 &     0.102 &    -0.005 &     0.000 &     0.000 &    -0.000\\
    & \pm 0.002 & \pm 0.000 & \pm 0.000 & \pm 0.000 & \pm 0.093 & \pm 0.001 & \pm 0.000 & \pm 0.000 & \pm 0.000 & \pm 0.000\\
  5 &    -1.513 &    -0.010 &     0.218 &     0.001 &   118.169 &     0.637 &    -0.178 &     0.002 &     0.012 &    -0.000\\
    & \pm 0.018 & \pm 0.000 & \pm 0.001 & \pm 0.000 & \pm 0.213 & \pm 0.012 & \pm 0.001 & \pm 0.000 & \pm 0.000 & \pm 0.000\\
 10 &    -3.110 &    -0.064 &     0.842 &     0.012 &   102.587 &     1.079 &    -0.665 &     0.005 &     0.080 &    -0.003\\
    & \pm 0.045 & \pm 0.000 & \pm 0.003 & \pm 0.000 & \pm 0.300 & \pm 0.029 & \pm 0.003 & \pm 0.001 & \pm 0.000 & \pm 0.000\\
 25 &    -6.498 &    -0.421 &     3.689 &     0.170 &    80.559 &     1.611 &    -2.757 &     0.037 &     0.553 &    -0.053\\
    & \pm 0.135 & \pm 0.001 & \pm 0.024 & \pm 0.000 & \pm 0.447 & \pm 0.087 & \pm 0.017 & \pm 0.011 & \pm 0.000 & \pm 0.000\\
 50 &    -9.905 &    -1.142 &     8.896 &     0.724 &    62.645 &     1.823 &    -6.351 &     0.285 &     1.614 &    -0.263\\
    & \pm 0.243 & \pm 0.004 & \pm 0.109 & \pm 0.002 & \pm 0.538 & \pm 0.182 & \pm 0.044 & \pm 0.050 & \pm 0.004 & \pm 0.001\\
100 &   -14.416 &    -2.291 &    17.091 &     2.204 &    43.088 &     2.096 &   -12.110 &     1.358 &     3.502 &    -0.978\\
    & \pm 0.256 & \pm 0.022 & \pm 0.311 & \pm 0.023 & \pm 0.512 & \pm 0.328 & \pm 0.080 & \pm 0.141 & \pm 0.031 & \pm 0.012\\
150 &   -18.046 &    -3.100 &    21.860 &     3.737 &    30.644 &     2.538 &   -16.358 &     2.590 &     4.835 &    -1.871\\
    & \pm 0.188 & \pm 0.057 & \pm 0.383 & \pm 0.079 & \pm 0.428 & \pm 0.377 & \pm 0.081 & \pm 0.165 & \pm 0.067 & \pm 0.032\\
200 &   -21.189 &    -3.751 &    24.196 &     5.191 &    21.244 &     3.107 &   -19.658 &     3.569 &     5.730 &    -2.803\\
    & \pm 0.218 & \pm 0.119 & \pm 0.315 & \pm 0.146 & \pm 0.392 & \pm 0.338 & \pm 0.074 & \pm 0.163 & \pm 0.079 & \pm 0.053\\
250 &   -23.885 &    -4.344 &    25.104 &     6.515 &    13.551 &     3.749 &   -22.330 &     4.194 &     6.324 &    -3.706\\
    & \pm 0.328 & \pm 0.189 & \pm 0.157 & \pm 0.186 & \pm 0.474 & \pm 0.255 & \pm 0.137 & \pm 0.193 & \pm 0.073 & \pm 0.075\\
300 &   -26.143 &    -4.920 &    25.273 &     7.673 &     6.966 &     4.418 &   -24.544 &     4.507 &     6.714 &    -4.552\\
    & \pm 0.457 & \pm 0.226 & \pm 0.302 & \pm 0.194 & \pm 0.695 & \pm 0.220 & \pm 0.263 & \pm 0.212 & \pm 0.118 & \pm 0.113\\
350 &   -27.966 &    -5.487 &    25.121 &     8.631 &     1.176 &     5.069 &   -26.380 &     4.593 &     6.963 &    -5.336\\
    & \pm 0.565 & \pm 0.212 & \pm 0.838 & \pm 0.242 & \pm 1.017 & \pm 0.381 & \pm 0.506 & \pm 0.270 & \pm 0.195 & \pm 0.179\\
 \end{tabular*}
 \end{ruledtabular}
\end{table*}

 To estimate the systematic uncertainties of the NN interaction at low
 energies we take different realistic potentials and compare their low
 energy threshold parameters.  Any estimate of the systematic errors
 based on variations of the potential form or possible radial
 dependences will provide a lower bound to the uncertainties. Besides
 the form of the potential used to fit the data, another source of
 systematic error is the selection of the data itself, due to addition
 of possible future data. Thus, the changes from the
 $3\sigma$-selected database of the Nijmegen analysis 20 years ago
 comprising $N=4301$ np and pp scattering data~\cite{Stoks:1993tb} to
 our recent $3\sigma$-self-consistent
 database~\cite{Perez:2013mwa,Perez:2013jpa} with about $N=6713$
 np and pp scattering data can be taken as an estimate on how much do
 we expect our predictions to change when a large body of new data is
 incorporated. 

Here we consider nine realistic local or minimally non-local
potentials (i.e. containing $L^2$ dependences or quadratic tensor
interactions) such as NijmII~\cite{Stoks:1994wp},
Reid93~\cite{Stoks:1994wp}, AV18~\cite{Wiringa:1994wb}, which provided
a $\chi^2/{\rm d.o.f} \sim 1$ to the Nijmegen
database~\cite{Stoks:1993tb}, and the new
DS-OPE~\cite{Perez:2013mwa,Perez:2013jpa},
DS-$\chi$TPE~\cite{Perez:2013oba,Perez:2013cza},
Gauss-OPE~\cite{Perez:2014yla}, Gauss-$\chi$TPE, DS-$\Delta$BO and
Gauss-$\Delta$BO which also provide a $\chi^2/{\rm d.o.f} \sim 1$ to
the Granada database~\cite{Perez:2013jpa}. To stress the obvious, we
associate the increase of about 2400 np and pp data from the Nijmegen
to the Granada databases with an additional systematic error,
foreseeing the possible impact that additional new data might have in
the future~\footnote{Note that the normality test foresees, within a
  confidence level, that when re-measurements of selected data are
  made, the statistical uncertainties will become smaller. It does not
  tell, however, what the error on interpolated energy or angle values
  would be. Thus new selected measurements will slightly change the
  most likely values for the parameters of the $\chi^2$-fit.}.  The
values of low energy parameters for the NijmII~\cite{Stoks:1994wp},
Reid93~\cite{Stoks:1994wp} have been determined
already~\cite{PavonValderrama:2005ku} but numerical precision has been
improved in the present work; the values for the remaining potentials
are also determined here. Our results are presented in
Table~\ref{tab:LEPS-statistic} (second line of each partial wave)
which shows the mean and standard deviation of the low energy
threshold parameters for the nine local potentials. Comparison of the
errors quoted in Table~\ref{tab:LEPS-statistic} clearly shows that the
main uncertainty in the low energy parameters is due to the different
representations or choices of high-quality potentials and not to the
propagation of experimental uncertainties used to fix the most-likely
potential chosen for the least squares $\chi^2$-analysis.

\begin{figure*}
\centering
\includegraphics[width=\linewidth]{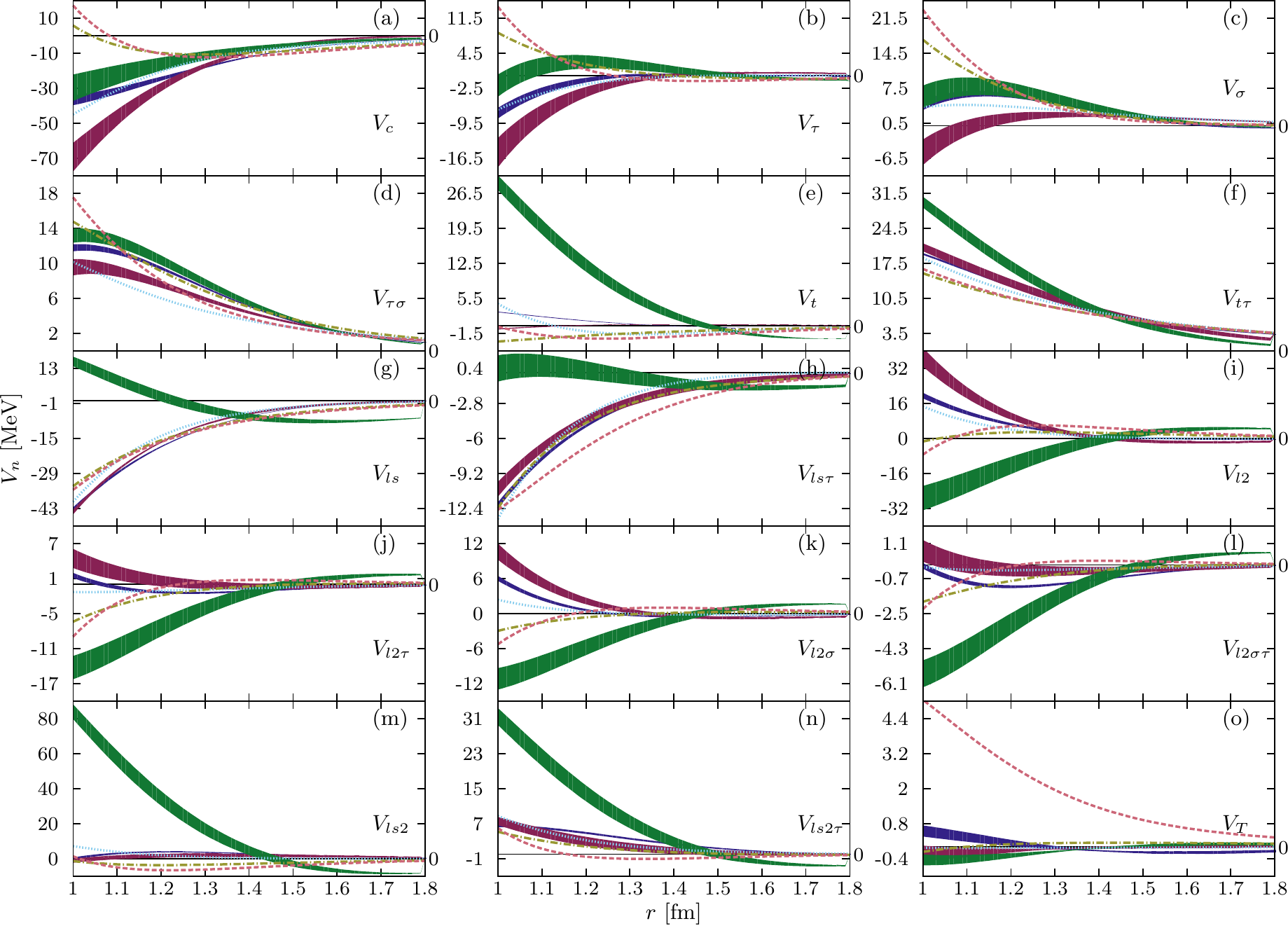}
\caption{Potentials in different channels in configuration space for
  short distances $1 < r < 1.8$fm. The blue, red and green bands
  represent the Gauss-OPE~\cite{Perez:2014yla}, Gauss-$\chi$TPE and
  Gauss-$\Delta$BO potentials respectively with their statistical
  uncertainty. For comparison we also show the local potentials
  AV18~\cite{Wiringa:1994wb} (dot-dashed olive green line),
  Reid93~\cite{Stoks:1994wp} (dotted light blue line) and
  NijmII~\cite{Stoks:1994wp} (dashed light red line).}
\label{fig:Voperators10_18}       
\end{figure*}

\begin{figure*}
\centering
\includegraphics[width=\linewidth]{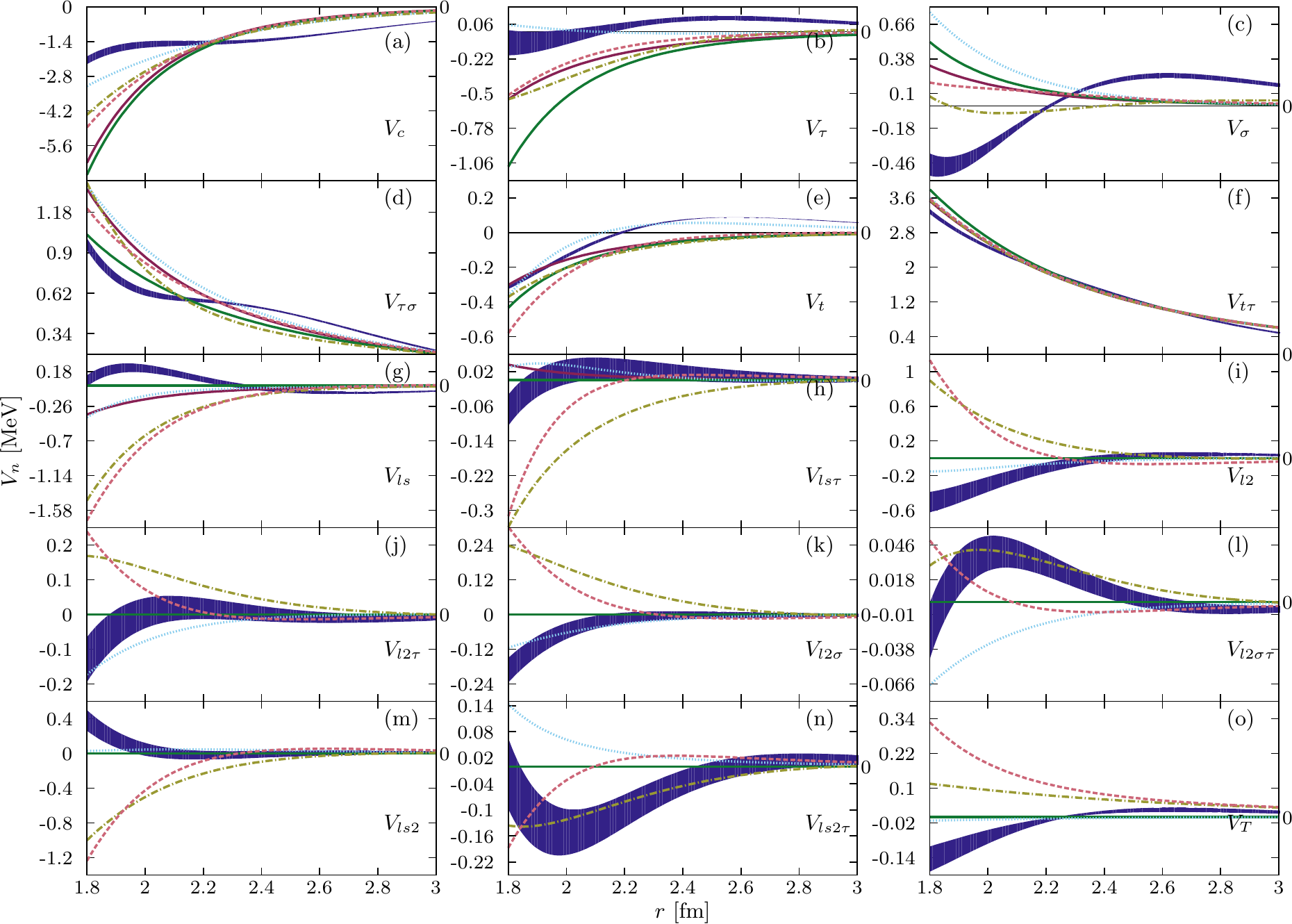}
\caption{Same as Fig.\ref{fig:Voperators10_18} for the intermediate
  range $1.8 < r < 3$fm.}
\label{fig:Voperators18_30}       
\end{figure*}

Our conclusions on systematic uncertainties at low energies are
vividly illustrated in Fig.~\ref{fig:v2v3v4}. As we see the spread of
all potential results is larger than the statistical error band, which
is our main point. In Figures \ref{fig:Voperators10_18} and
\ref{fig:Voperators18_30} we show our three local interactions
Gauss-OPE~\cite{Perez:2014yla}, Gauss-$\chi$TPE and Gauss-$\Delta$BO
in comparison with the AV18~\cite{Wiringa:1994wb},
Reid93~\cite{Stoks:1994wp} and NijmII~\cite{Stoks:1994wp} potentials
as a function of $r$. Again, it can clearly be seen that the spread of
the different interactions, which acounts for the systematic
uncertainty, is significantly larger than the statistical uncertainty
inferred in the potentials from the experimental errors of the
scattering data.  Notice however the different scale used on the
  $y$-axis of every pannel. This may make the discrepancies appear to
  be about the same order in all channels, which is not the
  case. Still, the systematic spread is always larger than the
  statistical error bands. 
An interpretation of the scattering data
and evaluation of the Skyrme coefficients or equivalent counterterms
in different partial waves was proposed in our previous
works~\cite{NavarroPerez:2012qr,Perez:2014kpa} with a focus on their
scale dependence and statistical uncertainties . The trend observed in
all physical observables in this paper confirms also the findings in
Ref.~\cite{Perez:2016vzj} regarding the systematic uncertainties.

This analysis is sufficient to prove that the lower bound on the
systematic error is at least one order of magnitude larger than the
statistical errors. Our results show that this conclusion is valid for
all the scattering properties, such as low energy parameters,
phase-shifts and amplitudes.

\section{On the portability of the partial wave analysis}
\label{sec:port}

One problem we want to address has to do with the portability of our
analysis and by extension of any PWA. One of the main technical
problems in carrying out a PWA in NN scattering is the inclusion of
many effects which are crucial to provide a convincing and
statistically sound fit to experimental data. Among them, the
inclusion of the long-range magnetic dipole local and anysotropic
interations needs summing up about 1000 partial waves and coordinate
space methods are strongly preferred over the momentum space
approaches, where implementation of these indispensable effects is a
real challenge still unsolved.

When this project was started the idea was to provide the most
relevant and portable information. Traditionally it has been thought
that phase-shifts with their corresponding covariance matrices
obtained from the fit represent the inherent uncertainty of the
interaction. Our experience does not support this view, and
statistically good fits to some phases at arbitrary energies do not
provide statistically good fits to scattering data, most often very
bad ones~\cite{Perez:2014bua}.  Another possibility which turns out to
provide a better approximation to our fit to data can be found by
fitting the Wolfenstein parameters within the systematic spread found
from the different potentials. This obviously incorporates the
correlations among the different phase shifts, but even if $\chi^2/\nu
\sim 1$ to the Wolfenstein parameters we typically find $\chi^2/\nu
\sim 4$ to the data.

However, in view of the results of the present paper where the form of
the potential itself representing the interaction provides the largest
uncertainty, we think that our results are best represented by an
average over the 6 potentials analyzed here. For the lower phases this
is summarized in Tables~\ref{tab:ppIsovectorPS},
\ref{tab:npIsovectorPS} and \ref{tab:npIsoscalarPS} and complete
tables can be provided upon request. As said, the corresponding errors
are comparable with the spread found using the previous PWA carried
out in the past.

\section{Conclusions and Outlook}
\label{sec:conclusions}

We summarize our results. In the present paper we have confronted
statistical vs systematic errors in the description of a largest body
of NN scattering below LAB energy $350 {\rm MeV}$ to date, namely
$6713$ np and pp scattering data collected from 1950 till 2013. We use
the classical statistical rules to evaluate the corresponding
uncertainties of the inferred potential, after the self-consistency of
the fit has been confidently established via checking Tail-Sensitive
normality tests~\cite{Perez:2014kpa}. We approach the determination of
systematic uncertainties by using 6 model potentials which
describe the same database comprising 6713 NN scattering data in a
statistically significant way. Thus, we have calculated and compared
phases and scattering amplitudes with their statistical
uncertainties. We have also calculated the low energy threshold
parameters of the coupled channel effective range expansion for all
partial waves with $j\leq 5$ by using a discrete version of the
variable S-matrix method. This approach provides satisfactory
numerical precision at a low computational cost, qualifying as a
suitable method to compare statistical vs systematic errors.
Statistical uncertainties are propagated via the bootstrap
method~\cite{Perez:2014jsa} where a family of DS-OPE potential
parameters is fitted after experimental data are replicated. We also
made a first estimate of the systematic uncertainties of the NN
interaction by taking nine different realistic potentials (i.e. with
$\chi^2/{\rm d.o.f} \lesssim 1$) and calculating the low energy
threshold parameters with each of them. These estimates should be
taken as a lower bound on the systematic uncertainties. In accordance
with preliminary estimates~\cite{NavarroPerez:2012vr, Perez:2012kt},
the systematic uncertainties tend to be at least an order of magnitude
larger than the statistical ones. The same trend between statistical
and systematic uncertainties is found when comparing phaseshifts and
scattering amplitudes. The present results encode the full PWA at low
energies and have a direct impact in {\it ab initio} nuclear structure
calculations in nuclear physics. The low energy threshold parameters
could also be used as a starting point to the determination and error
propagation of low energy interactions with the proper long distance
behavior in addition to the universal One Pion Exchange interaction.

Recently, there have been impressive bench-marking estimates on
uncertainties based on chiral NN and NNN forces in an order by order
scheme for light nuclei with $A \le 16$~\cite{Carlsson:2015vda}. Such
estimates are much larger than our simple preliminary estimates of
0.5 MeV in the binding energy per
nucleon~\cite{NavarroPerez:2012vr, Perez:2012kt}.  The upgrade of our
results incorporating the present systematic errors would increase our
estimate by a factor of four~\cite{Perez:2016vzj}.  The interactions
we have designed in this paper have the important property of being
statistically equivalent for a large NN database, and thus they can be
used to carry out comprehensive studies regarding different aspects of
binding in finite nuclei and in particular the impact on the
predictive power of nuclear structure and nuclear reactions
calculations which are now underway for the lighest $A=3,4$ systems
and will be analyzed in the future.


\acknowledgements

We thank the organizers and participants of the Workshop on Information and
Statistics in Nuclear Experiment and Theory (ISNET-3) taking place at
ECT* Trento, Italy for the lively and constructive discussions.

This work is supported by Spanish DGI (grant FIS2014-59386-P) and
Junta de Andaluc{\'{\i}a} (grant FQM225). This work was partly
performed under the auspicies of the U.S. Department of Energy by
Lawrence Livermore National Laboratory under Contract
No. DE-AC52-07NA27344.  Funding was also provided by the U.S.
Department of Energy, Office of Science, Office of Nuclear Physics
under Award No.  DE-SC0008511 (NUCLEI SciDAC Collaboration)

\appendix
\section{Model potentials}

We provide here details on the model potentials not described
previously. 

\subsection{The Gauss-$\chi$TPE Potential}
\label{app:GausTPE}

On this appendix we present the details of the Gaussian-$\chi$TPE
potential introduced on this article. The structure of the potential
is very similar to the Gaussian-OPE potential presented
in~\cite{Perez:2014yla}. The interaction is decomposed
as \begin{eqnarray} V(\vec r) = V_{\rm short} (r) \theta(r_c-r)+
  V_{\rm long} (r) \theta(r-r_c),
\label{eq:potential}
\end{eqnarray}
where  the short component is written as 
\begin{eqnarray}
   V_{\rm short}(\vec r) = \sum_{n=1}^{21} \hat O_n \left[\sum_{i=1}^N V_{i,n} F_{i} (r)\right] 
\label{eq:potential-short}
\end{eqnarray}
where $ \hat O_n$ are the set of operators in the extended AV18
basis~\cite{Wiringa:1994wb,NavarroPerez:2012vr,Perez:2012kt},
$V_{i,n}$ are unknown coefficients to be determined from data and
$F_{i}(r) = e^{-r^2/(2 a_i^2)}$ where $a_i = a/(1+i)$. $V_{\rm long}(\vec r)$
contains a Charge-Dependent (CD) One pion exchange (OPE) (with a
common $f^2=0.075$~\cite{NavarroPerez:2012vr,Perez:2012kt}), a charge
independent chiral two pion exchange ($\chi$-TPE) tail and electromagnetic (EM)
corrections which are kept fixed throughout. This corresponds to
\begin{eqnarray}
V_{\rm long}(\vec r) = V_{\rm OPE}(\vec r) + V_{\chi \rm TPE}(\vec r)  + V_{\rm em}(\vec r) \, . 
\end{eqnarray}
The boundary between the phenomenological $V_{\rm short}(r)$ and the
fixed $V_{\rm long}$ part $r_c$ is fixed at $1.8$fm. The parameter
$a$, that determines the width of each gaussian function was used as
an aditional fitting parameter obtaining the value $1.4335 \pm 0.0302$
fm. Table~\ref{tab:OperatorParameters} shows the values, with
statistical uncertainties, of the $V_{i,n}$ coefficients. Like in all
our previous potential analyses, although the form of the complete
potential is expressed in the operator basis the statistical analysis
is carried out more effectively in terms of some low and independent
partial waves contributions to the potential from which all other
higher partial waves are consistently deduced (see
Ref.~\cite{Perez:2013mwa,Perez:2013jpa}). For the chiral constants
determining $\chi$-TPE part we used the values obtained
on~\cite{Perez:2013oba} from fitting to the self-consistent NN
database, namely $c_1 =-0.41523 $, $c_3 = -4.66076$ and $c_4 =
4.31725$ GeV$^{-1}$. The resulting potential yields a merit figure of
$\chi^2/{\rm d.o.f.} = 1.09$ and the residuals tested positively for a
standard normal distribution after rescaling by the birge factor (see
Ref.~\cite{Perez:2014yla} and appendix~\ref{app:normaliy} here).

\begin{table*}[htb]
 \caption{\label{tab:OperatorParameters} Operator coefficients
   $V_{i,n}$ (in MeV) with their errors for the Gauss-$\chi$TPE,
   DS-$\Delta$BO and Gauss-$\Delta$BO potentials.  The coefficients of
   the $t T$, $\tau z$ and $\sigma \tau z$ operators are set to zero}
 \begin{ruledtabular}  
 \begin{tabular*}{\textwidth}{@{\extracolsep{\fill}}l D{.}{.}{3.4} D{.}{.}{3.4} D{.}{.}{4.4} D{.}{.}{3.4} D{.}{.}{3.4} D{.}{.}{4.4} D{.}{.}{3.4} D{.}{.}{3.4} D{.}{.}{4.4} }
   & \multicolumn{3}{c}{Gauss-$\chi$TPE} & \multicolumn{3}{c}{DS-$\Delta$BO} & \multicolumn{3}{c}{Gauss-$\Delta$BO} \\
   \cline{2-4} \cline{5-7} \cline{8-10}
   \noalign{\smallskip}
  $\hat{O}_n$  & \multicolumn{1}{c}{$V_1$} & 
          \multicolumn{1}{c}{$V_2$} & 
          \multicolumn{1}{c}{$V_3$} & 
          \multicolumn{1}{c}{$V_1$} & 
          \multicolumn{1}{c}{$V_2$} & 
          \multicolumn{1}{c}{$V_3$} &
          \multicolumn{1}{c}{$V_1$} & 
          \multicolumn{1}{c}{$V_2$} & 
          \multicolumn{1}{c}{$V_3$} \\
 \hline\noalign{\smallskip}
$c$             &      9.6837 &   -825.0435 &    945.4663 &     21.2098 &      4.2506 &     -2.3108 &     -7.3514 &   -157.2374 &     82.0322 \\
                & \pm  5.9084 & \pm 97.9106 & \pm 43.7258 & \pm  0.7545 & \pm  0.4744 & \pm  0.1302 & \pm  2.9646 & \pm 18.8009 & \pm 23.5960    \\
$\tau$          &     12.7705 &   -123.1648 &   -316.2405 &    -12.2427 &     -4.2044 &      0.7355 &    -17.2568 &    173.3607 &   -502.4170 \\
                & \pm  4.1353 & \pm 51.3913 & \pm 37.9668 & \pm  0.3630 & \pm  0.2142 & \pm  0.0460 & \pm  2.2635 & \pm 13.1983 & \pm 29.6577    \\
$\sigma$        &     18.5052 &    -33.2765 &   -421.9382 &    -25.8805 &     11.4536 &     -0.2993 &    -14.0544 &    211.4154 &   -524.6394 \\
                & \pm  3.9747 & \pm 37.6690 & \pm 42.1474 & \pm  0.4332 & \pm  0.1875 & \pm  0.0350 & \pm  2.7246 & \pm 10.9153 & \pm 19.4042    \\
$\tau \sigma$   &     24.6014 &     66.2296 &   -351.1114 &    -15.0644 &      9.4391 &      0.4022 &     -0.5228 &    167.3510 &   -351.4668 \\
                & \pm  4.3676 & \pm 19.7870 & \pm 16.5232 & \pm  0.1447 & \pm  0.1341 & \pm  0.0150 & \pm  2.5883 & \pm  6.1739 & \pm 12.4842    \\
$t$             &      4.2686 &    -23.5189 &     19.2503 &      0.0000 &      4.1612 &     -0.1994 &    -44.6418 &    340.6416 &   -295.9998 \\
                & \pm  1.1084 & \pm 10.5193 & \pm  9.5377 & \pm  0.0000 & \pm  0.1591 & \pm  0.0298 & \pm  1.2178 & \pm 18.4712 & \pm 17.8500    \\
$t \tau$        &     51.9125 &     23.5189 &    -75.4314 &      0.0000 &     21.2745 &      0.5108 &      0.8113 &    209.6308 &   -210.4421 \\
                & \pm  5.7086 & \pm 10.5193 & \pm  5.3862 & \pm  0.0000 & \pm  0.0938 & \pm  0.0226 & \pm  2.4346 & \pm  4.9184 & \pm  6.2049    \\
$ls$            &      2.9027 &   -401.1252 &    -33.0241 &    -22.4843 &    -14.0920 &     -1.1591 &    -93.5595 &    434.2099 &   -431.8457 \\
                & \pm  4.9308 & \pm 20.0357 & \pm  5.9716 & \pm  0.6346 & \pm  0.3326 & \pm  0.0545 & \pm  3.8004 & \pm 29.5342 & \pm 21.2608    \\
$ls \tau$       &     -3.7375 &    -76.3035 &    -34.3932 &     -7.4948 &     -4.0634 &     -0.2078 &    -17.1960 &     94.0333 &   -175.8844 \\
                & \pm  2.3106 & \pm 11.3094 & \pm  1.0782 & \pm  0.2115 & \pm  0.1696 & \pm  0.0348 & \pm  1.1839 & \pm  6.9873 & \pm  7.0953    \\
$l2$            &    -42.8284 &    494.8237 &    -97.5866 &     -3.1602 &     -7.4527 &      0.9505 &     73.0678 &   -506.6174 &    661.2658 \\
                & \pm  2.6104 & \pm 56.7825 & \pm  7.6448 & \pm  0.1346 & \pm  0.2228 & \pm  0.0459 & \pm  1.0393 & \pm 16.7099 & \pm 34.9700    \\
$l2 \tau$       &     -6.9591 &     45.6004 &     96.0890 &      2.1654 &     -5.7220 &      0.3142 &     29.7916 &   -242.6322 &    340.6513 \\
                & \pm  1.2133 & \pm 14.9971 & \pm  6.7830 & \pm  0.0623 & \pm  0.0995 & \pm  0.0172 & \pm  0.8288 & \pm  5.9885 & \pm 15.0560    \\
$l2 \sigma$     &    -15.0107 &    128.9282 &     90.3201 &      4.4383 &     -4.1874 &      0.3270 &     27.8688 &   -214.1951 &    312.4192 \\
                & \pm  1.1788 & \pm 19.0121 & \pm  9.1420 & \pm  0.0730 & \pm  0.0878 & \pm  0.0147 & \pm  0.6741 & \pm  5.0498 & \pm 13.3164    \\
$l2 \sigma \tau$&     -2.3215 &      0.9704 &     72.9793 &      2.5524 &     -1.7972 &      0.1081 &     11.5384 &    -97.7867 &    144.2162 \\
                & \pm  0.5389 & \pm  5.2064 & \pm  3.0161 & \pm  0.0243 & \pm  0.0487 & \pm  0.0054 & \pm  0.4459 & \pm  1.6616 & \pm  5.5114    \\
$ls2$           &     18.1617 &    -59.9055 &    -66.0483 &     -7.4948 &     20.8558 &     -1.4306 &   -140.3961 &   1021.2501 &   -863.6915 \\
                & \pm  1.8609 & \pm 21.5240 & \pm 11.9432 & \pm  0.2115 & \pm  0.3172 & \pm  0.0509 & \pm  2.5246 & \pm 49.6812 & \pm 42.5216    \\
$ls2 \tau$      &     -2.3108 &     88.2359 &    -68.7865 &     -2.4983 &      9.5656 &     -0.5098 &    -46.6431 &    375.7827 &   -351.7689 \\
                & \pm  0.9193 & \pm 13.2382 & \pm  2.1564 & \pm  0.0705 & \pm  0.1231 & \pm  0.0229 & \pm  1.1847 & \pm 14.0073 & \pm 14.1905    \\
$T$             &      0.4092 &     -6.9622 &     19.9535 &      0.8070 &      0.0731 &     -0.0759 &      1.8127 &    -13.8327 &     25.1923 \\
                & \pm  1.0118 & \pm  8.6713 & \pm 16.1725 & \pm  0.2166 & \pm  0.0401 & \pm  0.0159 & \pm  0.7598 & \pm  4.9259 & \pm  7.6163    \\
$\sigma T$      &     -0.4092 &      6.9622 &    -19.9535 &     -0.8070 &     -0.0731 &      0.0759 &     -1.8127 &     13.8327 &    -25.1923 \\
                & \pm  1.0118 & \pm  8.6713 & \pm 16.1725 & \pm  0.2166 & \pm  0.0401 & \pm  0.0159 & \pm  0.7598 & \pm  4.9259 & \pm  7.6163    \\
$l2 T$          &     -0.0682 &      1.1604 &     -3.3256 &     -0.1345 &     -0.0122 &      0.0126 &     -0.3021 &      2.3055 &     -4.1987 \\
                & \pm  0.1686 & \pm  1.4452 & \pm  2.6954 & \pm  0.0361 & \pm  0.0067 & \pm  0.0027 & \pm  0.1266 & \pm  0.8210 & \pm  1.2694    \\
$l2 \sigma T$   &      0.0682 &     -1.1604 &      3.3256 &      0.1345 &      0.0122 &     -0.0126 &      0.3021 &     -2.3055 &      4.1987 \\
                & \pm  0.1686 & \pm  1.4452 & \pm  2.6954 & \pm  0.0361 & \pm  0.0067 & \pm  0.0027 & \pm  0.1266 & \pm  0.8210 & \pm  1.2694    \\
 \end{tabular*}
 \end{ruledtabular}
\end{table*}

\subsection{The $\Delta$ Born-Oppenheimer potential}
\label{app:DeltaBO}

Here we will detail the Born-Oppenheimer potential with $\Delta$ terms
fitted to the Granada self-consistent database and introduced in this
work as a new source of systematic uncertainty of the NN interaction
(see e.g. \cite{RuizArriola:2009vp,Cordon:2011yd} for some
details). In general this potential has the same form of
Eq.(\ref{eq:potential}) with a clear boundary between the short range
phenomelogical part and the long-range pion-exchange tail; in
particular the long range part features squared Yukawa contributions
that result from including an intermediate $\Delta$ excitation.

The introduction of the $\Delta$-isobar as a dynamical degree of
freedom in the elastic NN channel requires extra terms on the
Lipmann-Schwinger equation. Applying the Born-Oppenheimer
approximation to second order results in more complicated structures
than just OPE and also a contribution to the central channel. This
additional terms are all proportional to $e^{-2 m_\pi r}$ in a
TPE-like fashion. The approximation can be expressed as
\begin{eqnarray}
 \bar{V}_{NN,NN}^{1\pi+2\pi+\dots}({\bm r}) &=&  V_{NN,NN}^{1\pi}({\bm r}) - 2 \frac{|V_{NN,N\Delta}^{1\pi}({\bm r})|^2}{\Delta} \nonumber \\
& - & \frac{1}{2} \frac{|V_{NN,\Delta\Delta}^{1\pi}({\bm r})|^2}{\Delta} + \mathcal{O}(V^3),
\label{eq:OPEDeltaBorn}
\end{eqnarray}
where $\Delta \equiv M_\Delta - M_N = 293$MeV and the transition potentials are
given by
\begin{eqnarray}
 V_{AB,CD} ({\bm r}) &=& (\bm{\tau}_{AB} \cdot {\bm \tau}_{CD})
 \left\{ {\bm \sigma}_{AB} \cdot {\bm \sigma}_{CD} \left[W_S^{1\pi}
   (r) \right]_{AB,CD} \right. \nonumber \\ 
   &+& \left. \left[S_{12} \right]_{AB,CD} \left[W_T^{1\pi}(r) \right]_{AB,CD}  \right\}, 
\end{eqnarray}
being $W_S^{1\pi}(r)$ and $W_T^{1\pi}(r)$ the usual spin-spin and
isovector tensor components of the one OPE potential. 

After dealing with the squared transitions of
Eq.(\ref{eq:OPEDeltaBorn}) and excluding the OPE part, the $\Delta$ Born-Oppenheimer potential
can be written as
\begin{eqnarray}
 V_{\Delta \rm BO}({\bm r}) &=& [V_C(r) + V_S(r){\bm \sigma}_1 \cdot {\bm \sigma}_2 + V_T(r)] \\
 &+&[W_C(r) + W_S(r){\bm \sigma}_1 \cdot {\bm \sigma}_2+ W_T(r)]  {\bm \tau}_1 \cdot {\bm \tau}_2, \nonumber
\end{eqnarray}
with components

\begin{widetext}
\begin{eqnarray}
 V_C(r) &=& - \frac{8 f_{\pi N\Delta}^2(f_{\pi N\Delta}^2 + 9 f_{\pi NN}^2)m_\pi^2}{81 \Delta}[2 Y_2(m_\pi r)^2 + Y_0(m_\pi r)^2], \nonumber \\
 V_S(r) &=& - \frac{4 f_{\pi N\Delta}^2(f_{\pi N\Delta}^2 -18 f_{\pi NN}^2)m_\pi^2}{243\Delta}[  Y_2(m_\pi r)^2 - Y_0(m_\pi r)^2], \nonumber \\
 V_T(r) &=&   \frac{4 f_{\pi N\Delta}^2(f_{\pi N\Delta}^2 -18 f_{\pi NN}^2)m_\pi^2}{243\Delta}[  Y_2(m_\pi r)^2 - Y_0(m_\pi r)Y_2(m_\pi r)], \nonumber \\
 W_C(r) &=&   \frac{4 f_{\pi N\Delta}^2(f_{\pi N\Delta}^2 -18 f_{\pi NN}^2)m_\pi^2}{243\Delta}[2 Y_2(m_\pi r)^2 + Y_0(m_\pi r)^2], \nonumber \\
 W_S(r) &=& + \frac{2 f_{\pi N\Delta}^2(f_{\pi N\Delta}^2 +36 f_{\pi NN}^2)m_\pi^2}{729\Delta}[  Y_2(m_\pi r)^2 - Y_0(m_\pi r)^2], \nonumber \\
 W_T(r) &=& - \frac{2 f_{\pi N\Delta}^2(f_{\pi N\Delta}^2 +36 f_{\pi NN}^2)m_\pi^2}{729\Delta}[  Y_2(m_\pi r)^2 - Y_0(m_\pi r)Y_2(m_\pi r)],
\end{eqnarray}
\end{widetext}
where $Y_0(x) = e^{-x}/x$ and $Y_2(x) = Y_0(x)(1+3/x+3/x^2)$. With
this form the coupling $f_{\pi N \Delta}$ can be determined by fitting
to NN data. The long range part is explicitly given by
\begin{eqnarray}
V_{\rm long}(\vec r) = V_{\rm OPE}(\vec r) + V_{\Delta \rm BO}(\vec r)  + V_{\rm em}(\vec r) \, . 
\end{eqnarray}

Note that we include the long range electromagnetic effects in
  the interaction and fits, but we only give the nuclear part of the
  phase-shifts.  See \cite{Perez:2013jpa} for details on how we deal
  with the electromagnetic interaction.

For the short range part given in Eq.(\ref{eq:potential-short}) we
implement two choices for the radial functions. The first one is a
Delta-Shell representation $F_i(r) = \Delta r \delta(r-r_i)$ with
$\Delta r= r_{i+1} - r_i = 0.6$fm and the second one is the guassian
functions given in the previous section. In both cases the cut radius
is set to $r_c = 1.8$fm. An initial fit was made to the Granada
self-consistent data base with the DS representation for the short
range part and including the $f_{\pi N \Delta}$ coupling as a fitting
parameter. A merit figure of $\chi^2/{\rm d.o.f.} = 1.12$ is
obtained. The potential parameters in the operator basis are given in
table~\ref{tab:OperatorParameters}; the fit gives a coupling of
$f_{\pi N \Delta} = (2.1778 \pm 0.0143) f_{\pi NN}$. A second fit uses
the SOG representation and the same fixed value of the $N \Delta$
coupling obtained with the previous fit. For this case the merit
figure is $\chi^2/{\rm d.o.f.} = 1.14$. The Potential parameters are
also given in table~\ref{tab:OperatorParameters}

\section{Normality of residuals}
\label{app:normaliy}

\begin{figure*}
\centering
\includegraphics[width=\linewidth]{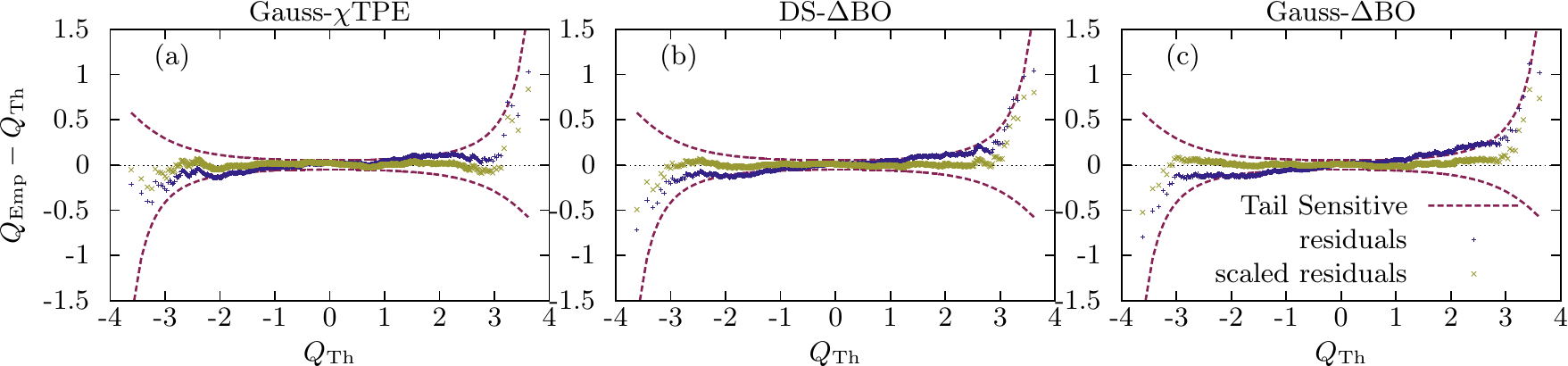}
\caption{Rotated Quantile-Quantile plot for the residuals (blue
  crosses) and scaled residuals (yellow diagonal crosses) of the three
  new phenomenological NN interactions presented in this work. The
   $95 \%$  confidence band of the Tail-Sensitive normality test are also given
  (red dashed lines).}
\label{fig:qqplots}       
\end{figure*}

It should be noted that the merit figure $\chi^2/{\rm d.o.f.}$ for the
potentials introduced on the previous appendices is slightly larger
than previous interactions fitted to the same database. In fact the
value in all three cases is outside of the $1\sigma$ confidence
interval $\chi^2/{\rm d.o.f.} = 1\pm \sqrt{2/{\rm d.o.f.}}$. This is a
consequence of the residuals, defined as
\begin{equation}
 R_i = \frac{O_i^{\rm exp}-O_i^{\rm theor}}{\Delta O_i},
\end{equation}
do not follow the standard normal distribution. However, a seemingly
unfavorable situation like this one can be salvaged by rescaling all
the residuals by a Birge factor defined as $B=1/\sqrt{\chi^2/{\rm
    d.o.f.}}$. The new merit figure will be, of course,
$\bar{\chi}^2/{\rm d.o.f.}  = B^2\chi^2/{\rm d.o.f.} = 1 $ by
definition. Nontheless, this rescaling does not guarantee that the
scaled residuals will follow the standard normal distribution; the
normality of the scaled residuals needs to be
tested. Figure~\ref{fig:qqplots} shows the rotated QQplot of the
residuals and scaled residuals for the three new potentials
Gauss-$\chi$TPE, DS-$\Delta$BO and Gauss-$\Delta$BO along with the
confidence bands of the particularly stringent Tail-Sensitive
normality test. In all three cases the original residuals do not
follow the stantard normal distribution, while the scaled ones do.

\section{The discrete variable-S-matrix method}
\label{sec:discrete-S}

The variable-$\mathbf{\hat{M}}(R,k)$
matrix equation is given by
\begin{eqnarray}
\frac{\partial \mathbf{\hat{M}}(R,k)}{\partial R} 
&=& \left( \mathbf{\hat{M}}(R,k) \mathbf{A}_k(R) - \mathbf{B}_k(R) \right) \mathbf{U}(R) \nonumber \\ &\times& \left( \mathbf{A}_k(R) \mathbf{\hat{M}}(R,k) - \mathbf{B}_k(R) \right),
\label{eq:variableM}
\end{eqnarray}
where  $R$ is the upper limit in the variable-phase equation, 
\begin{eqnarray}
\mathbf{A}_k(r) &=& {\rm diag} \left( \frac{\hat{j}_{l_1}(kr)}{k^{l_1+1}},\ldots,\frac{\hat{j}_{l_N}(k r)}{k^{l_N+1}} \right), \\
\mathbf{B}_k(r) &=& {\rm diag} \left( \hat{y}_{l_1}(kr) k^{l_1},\ldots,\hat{y}_{l_N}(kr)k^{l_N} \right) \, , 
\end{eqnarray}
 and $\mathbf{U}(R)$ is the reduced potential matrix. These are
 coupled non-linear differential equations which may become stiff in
 the presence of singularities in which case many integration points
 would be needed. For a discussion of these equations and their
 singularities in connection to the renormalization group and their
 fixed point structure see
 Refs.~\cite{PavonValderrama:2003np,PavonValderrama:2004nb,PavonValderrama:2007nu}.

For interactions which are smooth functions in configuration space
$\mathbf{U}(r)$, we propose a particular integration method by making
a delta-shell sampling of the interaction taking a sufficiently small
$\Delta r $. For simplicity we assume equidistant points $r_i = i \Delta r$
with $i=1, \dots , N$ and a maximum interaction radius $ r_{\rm max}=N \Delta r $  which corresponds to a  delta-shell representation
\begin{equation} 
\mathbf{\bar U}(r)=\sum_i \mathbf{U}(r_i)
\delta(r-r_i) \Delta r \, .   
\end{equation}
When substituting a DS potential, $\mathbf{U}(R)=\sum_i \Lambda_i
\delta(R-r_i)$, in Eq. (\ref{eq:variableM}) a recurrence relation is
obtained between the values of $ \mathbf{\hat{M}}$ on the left and
right side of each concentration radii $r_i$. That was the method used
in Ref.~\cite{PavonValderrama:2005ku}. In practice, it is numerically
better to solve the Schr\"odinger equation and matching logarithmic
derivatives piecewise as done in \cite{Perez:2013jpa}, yielding to
\begin{eqnarray}
 \mathbf{\hat{M}}(r_{i+\frac{1}{2}},k) &-& \mathbf{\hat{M}}(r_{i-\frac{1}{2}},k)  = \nonumber \\ 
& &  \left( \mathbf{\hat{M}}(r_{i+\frac{1}{2}},k) \mathbf{A}_k(r_i) - \mathbf{B}_k(r_i) \right)  \nonumber \\ &\times& \Lambda_i \left( \mathbf{A}_k(r_i) \mathbf{\hat{M}}(r_{i-\frac{1}{2}},k) - \mathbf{B}_k(r_i) \right).
\label{eq:recurrenceM}
\end{eqnarray}
Taking the low energy expansion in Eq. (\ref{MLowE}) and expanding also 
$\mathbf{A}_k$ and $\mathbf{B}_k$ 
\begin{eqnarray}
\mathbf{A}_k &=& \mathbf{A}_0 + \mathbf{A}_2 k^2 + \mathbf{A}_4 k^4 + \ldots \\
\mathbf{B}_k &=& \mathbf{B}_0 + \mathbf{B}_2 k^2 + \mathbf{B}_4 k^4 + \ldots
\end{eqnarray}
it is possible to obtain a recurrence relation for each matrix in
Eq. (\ref{MLowE}). The first two lowest terms in the expansion are
given by
\begin{eqnarray}
-\mathbf{a}^{-1}_{i+\frac{1}{2}} - \mathbf{a}^{-1}_{i-\frac{1}{2}} &=&
\left( \mathbf{a}^{-1}_{i+\frac{1}{2}} \mathbf{A}_0 + \mathbf{B}_0
\right) \nonumber \\
&\times& \Lambda_i \left(\mathbf{A}_0 \mathbf{a}^{-1}_{i-\frac{1}{2}} +
\mathbf{B}_0 \right) \label{eq:rec_alfa} \\
\mathbf{r}_{i+\frac{1}{2}} - \mathbf{r}_{i-\frac{1}{2}} &=& -2\left( \mathbf{a}^{-1}_{i+\frac{1}{2}} \mathbf{A}_0 + \mathbf{B}_0 \right) \nonumber \\
&\times& \Lambda_i \left(\frac{1}{2} \mathbf{A}_0 \mathbf{r}_{i-\frac{1}{2}} -\mathbf{A}_2 \mathbf{a}^{-1}_{i-\frac{1}{2}} - \mathbf{B}_2 \right) \nonumber \\
&-& \left(\frac{1}{2} \mathbf{r}_{i+\frac{1}{2}} \mathbf{A}_0 -\mathbf{a}^{-1}_{i+\frac{1}{2}} \mathbf{A}_2 - \mathbf{B}_2 \right) \nonumber \\
&\times& \Lambda_i \left( \mathbf{A}_0 \mathbf{a}^{-1}_{i-\frac{1}{2}} - \mathbf{B}_0 \right) \, .\label{eq:rec_reff}
\end{eqnarray}
Higher orders can straightforwardly be written, but the final
formulas are rather long and will not be quoted here.  Note the
hierarchy of the equations where low energy parameters to a given
order involve the same or lower orders only.  These recursive
equations are reversible, i.e. going upwards or downwards are inverse
operations of each other on the discrete radial grid. They appeared
for S-waves in Ref.~\cite{Entem:2007jg} for the scattering length and
the effective range. The initial condition corresponds to taking a
trivial solution,
\begin{eqnarray}
\mathbf{a}_{-\frac{1}{2}} = \mathbf{r}_{-\frac{1}{2}} = \dots =0
\end{eqnarray}
whereas the final value provides the sought low energy parameters 
\begin{eqnarray}
\mathbf{a}= \mathbf{a}_{N+\frac{1}{2}} \, , \qquad \mathbf{r}= \mathbf{r}_{N+\frac{1}{2}} \, , \qquad \dots 
\end{eqnarray}
A good feature of these discretized variable phase-like equations is
that they jump over singularities. The calculation of the low energy
threshold parameters with a DS potential is very similar to the
calculation of phase-shifts detailed in the appendix B
of~\cite{Perez:2013jpa} and is also the discrete analogous of the
variable S matrix  method of~\cite{PavonValderrama:2005ku}.


\end{document}